\documentclass[useAMS,usenatbib]{mn2e}
\usepackage{hyperref}

\usepackage{graphicx}

\usepackage{longtable}
\usepackage{tabularx}
\usepackage[flushleft]{threeparttable}
\usepackage{lscape}
\usepackage{amssymb}

\title[The WISE View of RV Tauri Stars]{The WISE View of RV Tauri Stars}

\author[I. Gezer et al.]
   {I.~Gezer$^{1,2}$,
   H.~Van Winckel$^{2}$, Z.~Bozkurt$^{1}$, K.~De Smedt$^{2}$, D.~Kamath$^{2}$,
   \newauthor
   M.~Hillen$^{2}$ and R.~Manick$^{2}$ \\
  $^{1}$Astronomy and Space Science Department, Ege University ,35100 Bornova, Izmir, Turkey\\
  $^{2}$Institute of Astronomy, KU Leuven, Celestijnenlaan, 200D 3001 Leuven, Belgium}
\begin{document}


\pagerange{\pageref{firstpage}--\pageref{lastpage}} \pubyear{2002}

\maketitle

\label{firstpage}

\begin{abstract}
  We present a detailed study based on infrared photometry of all
  Galactic RV Tauri stars from the General Catalogue of Variable Stars
  (GCVS). RV Tauri stars are the brightest among the population II
  Cepheids. They are thought to evolve away from the asymptotic giant
  branch (AGB) towards the white dwarf
  domain.  IRAS detected several RV Tauri stars because of their large
  IR excesses and it was found that they occupy a specific region in
  the [12] $-$ [25], [25] $-$ [60] IRAS two-colour diagram. We used
  the all sky survey of WISE to extend these studies and compare the
  infrared properties of all RV\,Tauri stars in the GCVS with a 
  selected sample of post-AGB objects with the goal to place the
  RV\,Tauri pulsators in the context of post-AGB evolution. Moreover,
  we correlated the IR properties of both the RV\,Tauri stars and the
  comparison sample with other observables like binarity and
  the presence of a photospheric chemical anomaly called depletion.
  We find that Galactic RV Tauri stars display a range of infrared
  properties and we differentiate between disc sources, objects with
  no IR excess and objects for which the spectral energy distribution
  (SED) is uncertain. We obtain a clear  correlation between disc sources and binarity.
  RV Tauri stars with a variable mean magnitude are exclusively found among the disc
  sources. We also find evidence for
  disc evolution among the binaries. Furthermore our studies show that the presence of
  a disc seems to be a necessary but not sufficient condition for the
  depletion process to become efficient.

\end{abstract}


\section{Introduction}

RV Tauri stars are luminous population II Cepheid variables with
spectral types typically between F and K. 
They are named after the prototype RV Tau. According to the GCVS \citep{samus09}, there are 153 RV\,Tauri variables in our
Galaxy and several have been found in the LMC and SMC
\citep{alcock98,soszynski08,buchler09,soszynski10}. 
In this paper, we focus on the Galactic sample. 

Characteristic and defining light curves of RV Tauri stars show subsequent deep and 
shallow minima \citep[e.g.][]{pollard97}. Their formal pulsation period is in between 30 and 150 days and the amplitude may reach up to 4 magnitudes in {$\it{V}$}. The pulsations of RV\,Tauri stars can be regular, however, irregular pulsations are more commonly
detected. There are several explanations for the light variations
\citep{tuchman93}. Non-linear, non-adiabatic
hydrodynamical RV Tauri models \citep{fokin94} support the hypothesis in which a
resonance between the fundamental mode and the first overtone is
responsible for the alternating minima in the light curves.

RV Tauri stars are divided into two photometric subclasses based on
the characteristics of the light curve \citep{kukarkin58}. These
subclasses are indicated with letter $\bf{a}$ and $\bf{b}$. The RVa stars 
have a constant mean magnitude, while the RVb stars have a varying
mean magnitude on a time scale between 600 and 1500
days \citep{samus09}. In addition to the photometric subclasses, RV Tauri stars are divided into 
three spectroscopic subclasses \citep{preston63}. These subclasses are
labelled as $\bf{A}$, $\bf{B}$, $\bf{C}$. The stars of type RVA are of spectral type G-K, and show strong absorption lines and normal CN or CH bands while TiO bands
sometimes appear at photometric minimum. RVB stars are generally
somewhat hotter, weaker lined objects which show enhanced CN and CH
bands. RVC stars are also weak
lined but they show normal bands of CH and CN. This last subclass
contains the globular cluster members of the RV Tauri class. There is no
correlation between the spectroscopic and photometric classes.

Earlier pioneering ground-based detections of
IR excesses in some RV Tauri stars (AC Her, U Mon,
R Sct, R Sge) \citep{gehrz70,gehrz72a,gehrz72b} were followed by systematic full-sky
surveys once  IR satellites were launched.
A limited number of RV Tauri stars were
detected by IRAS (Infrared Astronomical Satellite 1983) confirming the ground-based detections of some, 
and enlarging the sample for which a large IR excess due to the thermal emission from
circumstellar dust was found. This was used to classify RV Tauri stars
as post-Asymptotic Giant Branch (post-AGB) stars by \cite{jura86}.
Assuming that the RV Tauri stars are pulsating
post-AGB stars which are crossing the instability strip, one may
expect that the photospheres are enriched by AGB nucleosynthesis
products. Indeed, a rich nucleosynthesis occurs on the AGB
evolutionary phase, where carbon, nitrogen, and the slow neutron
capture ({$\it{s}$}-process) elements beyond iron
are produced. Recurrent mixing episodes, induced by the
thermal pulses on the AGB transport the
freshly synthesized material to the surface \citep[e.g.][]{herwig05}.
Therefore, abundances of Galactic RV Tauri stars have been 
extensively studied over the past two decades. These
studies show that RV Tauri stars are mainly {$\it{not}$} enriched 
in C and s-process elements \citep[e.g.][]{giridhar94,giridhar98,giridhar00,
gonzalez97b,gonzalez97a, vanwinckel98, maas02,maas03,maas05,deroo05a}. 
Earlier work suggested, for instance, the two prototypical
RV Tauri stars AC Her and RU Cen to be C-enriched, based on their high [C/Fe]
\citep{gehrz70,gehrz72a}. Also the dust component in these two objects was
presumed to be C-rich. Both assertions are, however, no longer supported by
more recent works, like the photospheric abundance analysis of \cite{vanwinckel98}
and the dust studies of \cite{molster02a}, \cite{molster02b}, \cite{molster02c},
\cite{gielen07,gielen09b, hillen15}. One exception may be V453\, Oph, which
shows a mild {$\it{s}$}-process overabundance but
it is not accompanied by C enhancement \citep{deroo05a}. 
Also the abundance analysis results on a LMC sample of RV Tauri
stars \citep{reyniers07a,reyniers07b} showed that C and
{$\it{s}$}-process enrichments are generally not found with one exception
being MACHO 47.2496.8 which shows a strongly enriched photosphere
\citep{reyniers07a}.

While C and {$\it{s}$}-process enrichments are generally not found, many 
RV\,Tauri photospheres show a distinct chemical anomaly which is called
\textquotedblleft depletion\textquotedblright. Depleted photospheres display a similar chemical
composition as the gas phase of the interstellar medium: refractory
elements, which have high dust condensation temperature, are underabundant,
while volatiles, which have low condensation temperature, are more
abundant \citep[e.g.][]{vanwinckel03}. The process to acquire the chemical anomaly is not
completely understood yet, but it must be a chemical rather than a
nucleosynthetic process. Dust formation may lead to a chemical
fractionation event in the circumstellar environment. To obtain a
depleted photosphere, the radiation pressure on circumstellar dust
grains must fractionate the dust from the gas. The cleaned gas is then
re-accreted onto the stellar surface. As a result of this process,
stars display peculiar photospheric composition similar to the
interstellar gas.

The recurrent chemical anomalies found in RV\,Tauri stars 
make that the evolutionary nature of these pulsators are still not
clear. In this contribution we focus on the dust
excesses. \cite{evans85} showed that infrared colours of RV Tauri stars occupy a
specific region in the IRAS [12]-[25], [25]-[60] diagram (see Fig.~1)
and he defined this region as the {$\it{RV}$} {$\it{Tauri}$}
{$\it{Box}$}. Unfortunately, IRAS colours are available for only 18 RV
Tauri stars.

Here, we present a systematic study of the IR properties of RV Tauri
stars and we use all Galactic RV\,Tauri stars listed in the GCVS
\citep{samus09} as our sample.  The IR photometry is a good tracer of
the temperature gradient of the circumstellar dust (if present), which
we want to correlate with the properties of the central object. We
expand the RV Tauri study based on IRAS colours to a similar but much
deeper one using WISE (Wide-field Infrared Survey Explorer) data. The
WISE satellite was launched in December 2009 and scanned the whole sky
in 3.4, 4.6, 12 and 22 $\mu$m bands \citep{wright10}.  The biggest
advantage is that the WISE survey is deep enough to detect all
Galactic RV Tauri stars of our sample.

The WISE photometric bands are not the same as the IRAS bands, so we
defined a new colour combination [3.4]$-$[4.6] (W1-W2) versus
[12]$-$[22] (W3-W4) as a good WISE alternative, be it more focussed on
the near- and mid-IR. The
[3.4]$-$[4.6] colour is a good indicator for the presence of a near IR
excess while the [12]$-$[22] colour is indicative of the presence of
cooler dust. In order to compare the Galactic RV Tauri stars with 
post-AGB objects, we compose a reference sample, which consist
of 60 well-known dusty optically bright post-AGB stars. 

Our samples and the WISE colour-colour diagram are presented in
Sect. 2. In Section 3 we study the correlation between the colours and other 
observational properties of RV Tauri stars and compare the RV Tauri sample 
with the well defined post-AGB stars. The relation between IR excess, 
binarity and chemical anomalies are discussed in section 4. Conclusions are summarized in Sect. 5.

\begin{figure}
\begin{center}
\includegraphics[bb= 0 0 1200 800, width=9cm,angle=0]{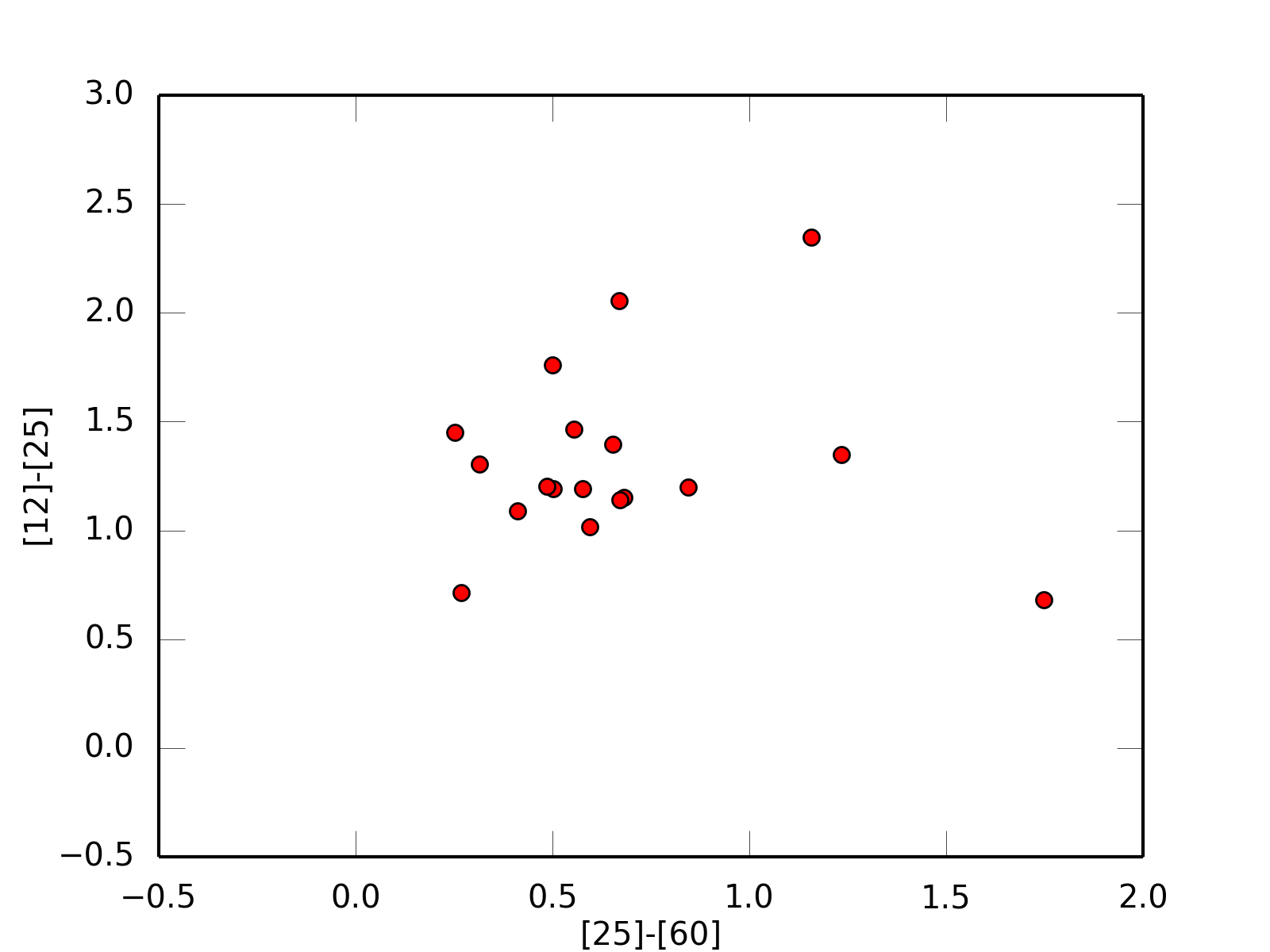}
\caption{IRAS colour-colour diagram for RV Tauri stars, RV Tau box \citep{evans85}.}  
\end{center}
\label{RVTauBox}
\end{figure}

\begin{figure}
\begin{center}
\begin{tabular}{c}
\resizebox{\hsize}{!}{\includegraphics[bb= 0 0 1200 800, width=7cm,angle=0]{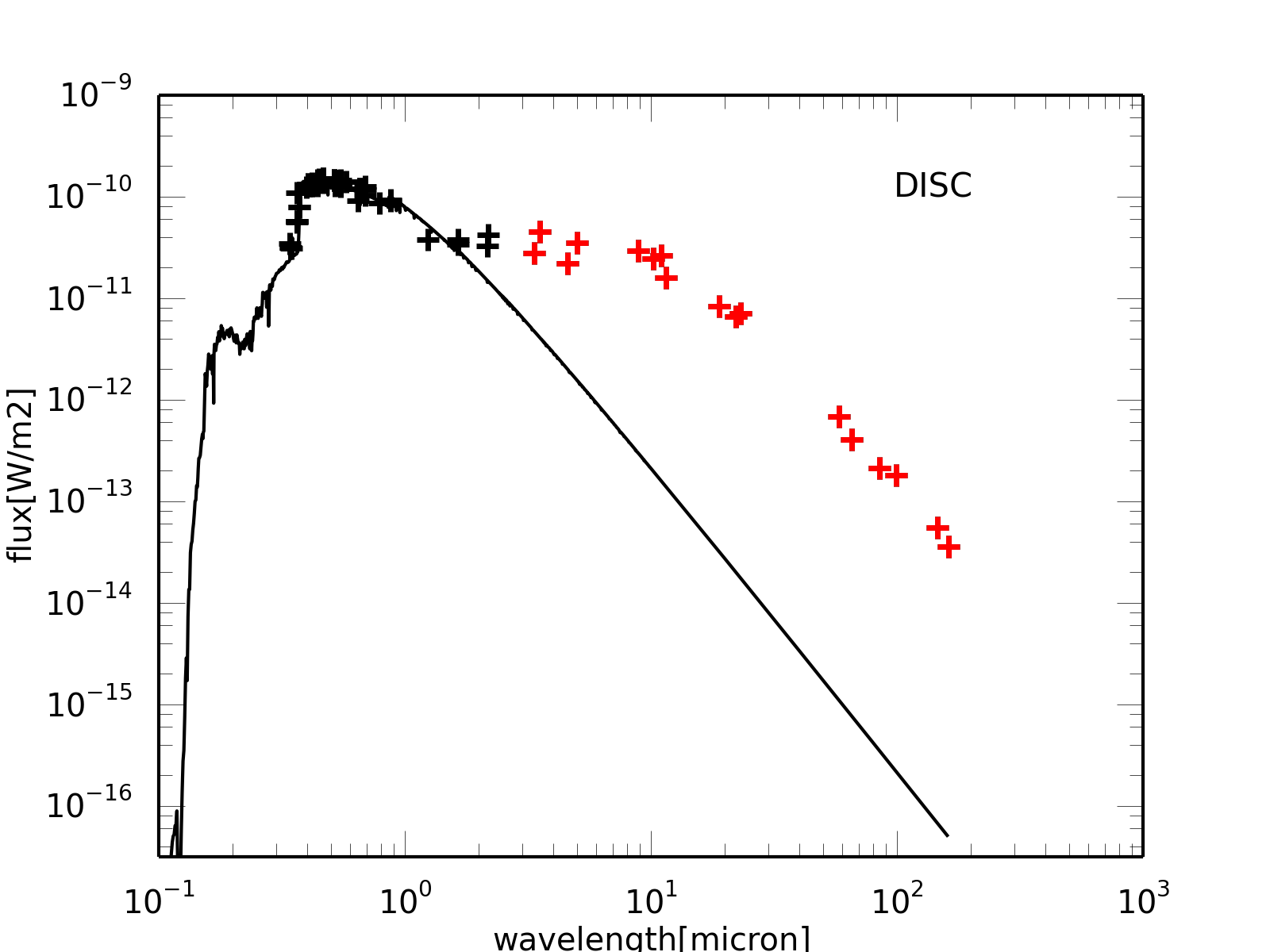}}\\
\\
\resizebox{\hsize}{!}{\includegraphics[bb= 0 0 1200 800, width=7cm,angle=0]{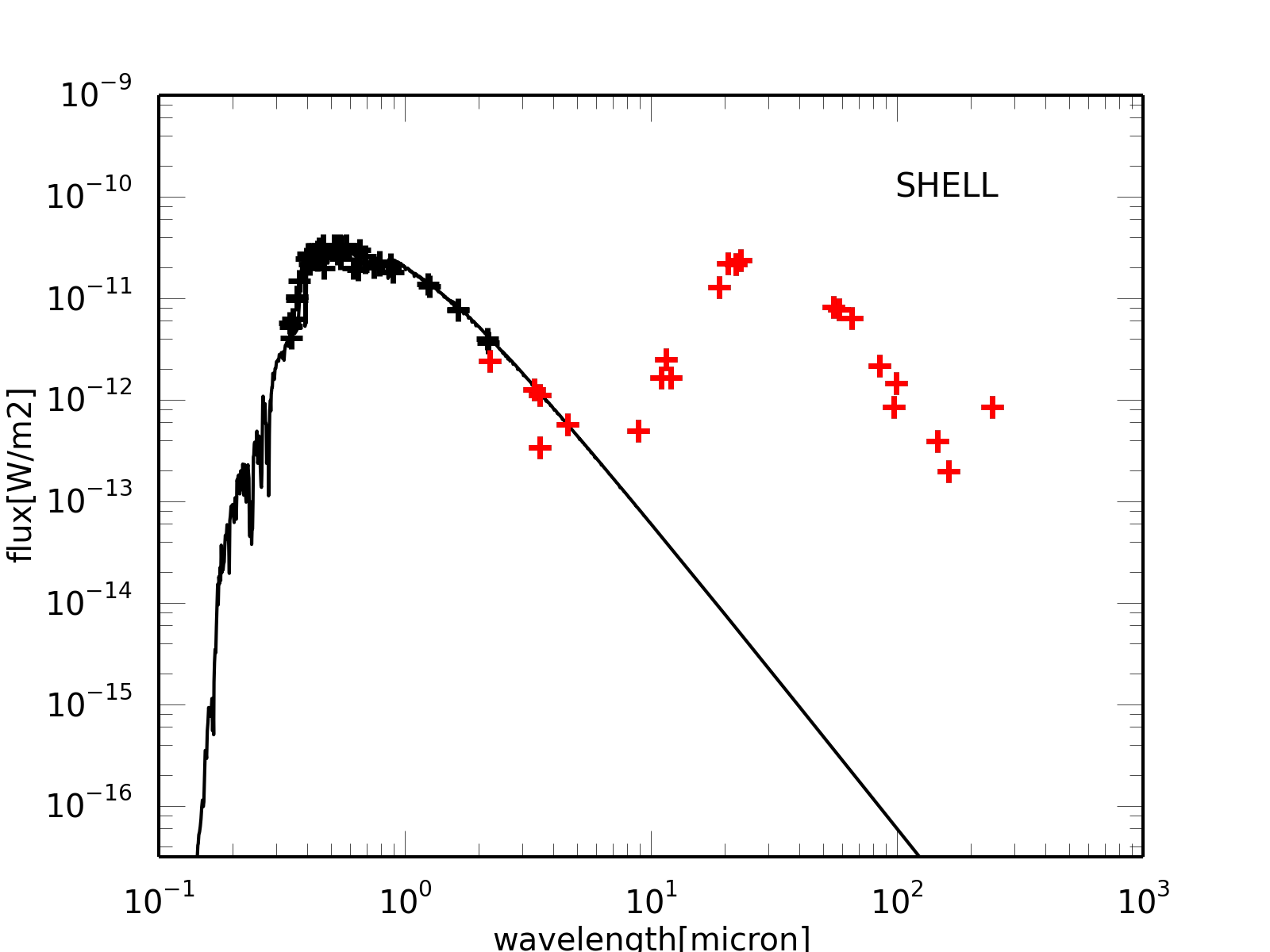}}\\
\end{tabular}{c}
\caption{Characteristic SEDs of optically bright post-AGB stars. Top
  panel: the SED of  89 Her which is a well-known binary system. The
  solid line represents the photospheric flux of the model atmosphere,
  the black crosses the dereddened photometric data.
The red plusses show the IR photometric data, pointing to the presence of a
compact disc \citep{hillen14}. The bottom panel shows the SED of
HD\,161796, the solid line represents the photosphere, the black
crosses the dereddened measurements, while the red
crosses show the IR data, pointing to the presence of an expanding
dusty shell \citep{min13}.}
\end{center}
\label{discshell}
\end{figure}

\section{Reference Sample}

Since RV Tauri stars are a subset of post-AGB objects we compare the
Galactic RV Tauri stars with Galactic post-AGB objects that do not
display the RV\,Tauri light curves.  To do this we have defined a
reference sample of well characterised Galactic post-AGB stars.
Optically bright post-AGB stars show typically two different types of
SED (see Fig.~2) \citep{vanwinckel03}.  The first group of post-AGB
stars display SEDs with a broad IR excess starting in the near-IR
region,  which points to the presence of hot dust near the star.  The
[3.4]-[4.6] colour is a good identifier for these sources which are
called {$\it{disc}$} sources (see section 2.1). The second group
contains post-AGB stars with a double peaked SED (see Fig.~2) where
the first peak corresponds to the photospheric flux and the second
peak represents the IR emission of an expanding dusty envelope. In the
outflow model for a single post-AGB star, the near-infrared excess is
expected to disappear within years after the cessation of the dusty
mass loss. The mid-infrared excess is an indicator of this detached
expanding circumstellar envelope (CSE). The [12]-[22] colour is a good
identifier for these sources which are called {$\it{shell}$} sources
(see section 2.2).

\begin{table*}
 \centering
  \caption{Reference sample, consisting of well selected and observationally characterized post-AGB objects.}
  \begin{tabular}{llllccccc}
  \hline
   IRAS          & Other Name & RA& DEC  & Spectral Type    & [12]$-$[22] & [3.4]$-$[4.6]& SED & Ref  \\
  \hline

01427$+$4633 &BD$+$46 442 &   01 45 47.03&  $+$46 49 00.97   &    F2III 	 & 1.434 & 0.83   & disc &  2	  \\
04296$+$3429 &		  &   04 32 56.97&  $+$34 36 12.40   &    G0Ia  	 & 2.536 & 1.055  & shell&  12,7  \\
05040$+$4820 &BD$+$48 1220&   05 07 50.30&  $+$48 24 09.42   &    A4Ia  	 & 6.095 & 0.081  & shell&  7,8   \\
05089$+$0459 &		  &   05 11 36.15&  $+$05 03 26.30   &    M3I		 & 2.511 & 0.875  & shell&  7	  \\
05113$+$1347 &		  &   05 14 07.76&  $+$13 50 28.30   &    G8Ia  	 & 2.908 & 0.514  & shell&  7,16  \\
05208$-$2035 &		  &   05 22 59.42&  $-$20 32 53.03   &    Me... 	 & 1.293 & 0.824  & disc &  1	  \\
05238$-$0626 &BD$-$06 1178&   05 26 19.76&  $-$06 23 57.40   &    ~		 & 2.468 & 0.291  & shell&  7	  \\
05341$+$0852 &		  &   05 36 55.05&  $+$08 54 08.68   &    F5I		 & 2.225 & 1.065  & shell&  12,7  \\
05381$+$1012 &HD 246299   &   05 40 57.05&  $+$10 14 24.99   &    G2/G3I	 & 2.671 & 0.34   & shell&  7	  \\
06160$-$1701 &UY CMa	  &   06 18 16.37&  $-$17 02 34.72   &    G0V		 & 1.24  & 1.33   & disc &  1	  \\
06176$-$1036*&HD 44179	  &   06 19 58.22&  $-$10 38 14.71   &    B8V		 & ---   & ---    & disc &  1	  \\
06338$+$5333 &HD 46703	  &   06 37 52.43&  $+$53 31 01.96   &    F7IVw 	 & 0.96  & 0.271  & disc &  1	  \\
06472$-$3713 &ST Pup	  &   06 48 56.41&  $-$37 16 33.34   &    G2I		 & 1.625 & 1.31   & disc &  1	  \\
06530$-$0213 &		  &   06 55 31.82&  $-$02 17 28.30   &    G1I		 & 2.643 & 0.424  & shell&  6	  \\
07008$+$1050 &HD 52961	  &   07 03 39.63&  $+$10 46 13.06   &    A0		 & 0.922 & 0.985  & disc &  1	  \\
07134$+$1005 &HD 56126	  &   07 16 10.26&  $+$09 59 47.99   &    F5Iab 	 & 2.527 & 0.137  & shell&  11,12 \\
07140$-$2321 &SAO 173329  &   07 16 8.27 &  $-$23 27 01.61   &    F5		 & 1.393 & 1.128  & disc &  1	  \\
07171$+$1823 &		  &   07 20 1.53 &  $+$18 17 26.12   &    BQ[]  	 & 3.15  & 0.955  & shell&  7	  \\
07430$+$1115 &		  &   07 45 51.39&  $+$11 08 19.60   &    G5Ia  	 & 2.544 & 0.646  & shell&  7	  \\
08187$-$1905 &HD 70379	  &   08 20 57.10&  $-$19 15 03.47   &    F6Ib/II	 & 4.551 & 0.069  & shell&  6	  \\
09060$-$2807 &		  &   09 08 10.13&  $-$28 19 10.39   &    F5		 & 1.067 & 1.461  & disc &  1	  \\
09371$+$1212*&Frosty Leo  &   09 39 53.96&  $+$11 58 52.60   &    M4		 & ---   & ---    & shell&  19	  \\
10158$-$2844*&HR 4049	  &   10 18 7.59 &  $-$28 59 31.20   &    B9.5Ib-II	 &   --- & ---     & disc &  1	  \\
10174$-$5704 &		  &   10 19 16.89&  $-$57 19 26.00   &    G8IaO 	 & 1.672 & 1.192  & disc &  1	  \\
10456$-$5712*&		  &   10 47 38.40&  $-$57 28 02.68   &    M1II  	 & ---   & ---    & disc &  1	  \\
11000$-$6153 &HD 95767	  &   11 02 4.31 &  $-$62 09 42.84   &    F2III 	 & 0.965 & 1.689  & disc &  1	  \\
11472$-$0800 &		  &   11 49 48.04&  $-$08 17 20.47   &    F5Iab:	 & 1.798 & 1.723  & disc &  1	  \\
12222$-$4652 &HD 108015   &   12 24 53.50&  $-$47 09 07.51   &    F4Ib/II	 & 1.157 & 1.797  & disc &  1	  \\
13258$-$8103 &		  &   13 31 7.05 &  $-$81 18 30.40   &    F4Ib-G0Ib	 & 2.022 & 1.272  & disc &  1	  \\
14488$-$5405 &		  &   14 52 28.72&  $-$54 17 42.80   &    ~		 & 3.468 & 0.791  & shell&  18	  \\
14524$-$6838 &EN TrA	  &   14 57 00.68&  $-$68 50 22.89   &    F2Ib  	 & 1.097 & 1.592  & disc &  1	  \\
15039$-$4806 &HD 133656   &   15 07 27.44&  $-$48 17 53.87   &    A1/A2Ib/II	 & 4.763 & 0.048  & shell&  5	  \\
15469$-$5311 &		  &   15 50 43.80&  $-$53 20 43.33   &    F3		 & 1.449 & 1.624  & disc &  1	  \\
15556$-$5444 &		  &   15 59 32.57&  $-$54 53 20.40   &    F8		 & 1.212 & 1.929  & disc &  1	  \\
16230$-$3410 &		  &   16 26 20.39&  $-$34 17 12.76   &    F8		 & 1.314 & 1.48   & disc &  1	  \\
17038$-$4815**&		  &   17 07 36.64&  $-$48 19 08.56   &    G2p...	 & 1.199 & 1.501  & disc &  1	  \\
17074$-$1845 &BD$-$18 4436&   17 10 24.15&  $-$18 49 00.67   &    B5Ibe 	 & 4.448 & 0.15   & shell&  13	  \\
17233$-$4330 &		  &   17 26 58.65&  $-$43 33 13.47   &    G0p...	 & 1.423 & 1.834  & disc &  1	  \\
17243$-$4348 &LR Sco	  &   17 27 53.62&  $-$43 50 46.27   &    G2		 & 1.231 & 1.715  & disc &  1	  \\
17436$+$5003 &HD 161796   &   17 44 55.47&  $+$50 02 39.48   &    F3Ib  	 & 4.418 & 0.127  & shell&  10	  \\
17534$+$2603 &89 Her	  &   17 55 25.19&  $+$26 02 59.97   &    F2Ibe 	 & 1.106 & 0.728  & disc &  1	  \\
18095$+$2704 &V887 Her	  &   18 11 30.67&  $+$27 05 15.61   &    F3Ib  	 & 2.322 & 0.773  & shell&  17	  \\
18123$+$0511 &		  &   18 14 49.39&  $+$05 12 55.70   &    G5		 & 1.471 & 1.765  & disc &  1	  \\
18158$-$3445 &		  &   18 19 13.37&  $-$34 44 30.00   &    F6		 & 1.59  & 1.379  & disc &  1	  \\
18379$-$1707 &		  &   18 40 48.62&  $-$17 04 38.30   &    B1IIIpe	 & 4.111 & 0.672  & shell&  13	  \\
19114$+$0002 &HD 179821   &   19 13 58.61&  $+$00 07 31.93   &    G5Ia  	 & 3.676 & 0.629  & shell&  14	  \\
19135$+$3937 &		  &   19 15 12.14&  $+$39 42 50.51   &    ~		 & 1.306 & 1.214  & disc &  2	  \\
19157$-$0247 &		  &   19 18 22.71&  $-$02 42 10.89   &    B1III 	 & 1.221 & 1.713  & disc &  1	  \\
19410$+$3733 &HD 186438   &   19 42 52.92&  $+$37 40 41.46   &    F3Ib  	 & 0.915 & 1.395  & disc &  4	  \\
19475$+$3119 &HD 331319   &   19 49 29.56&  $+$31 27 16.22   &    F8		 & 5.242 & 0.093  & shell&  9	  \\
19500$-$1709 &HD 187885   &   19 52 52.70&  $-$17 01 50.30   &    F2/F3Iab+...   & 2.564 & 0.367  & shell&  12,15 \\
19580$-$3038 &V1711 Sgr   &   20 01 7.98 &  $-$30 30 38.89   &    F5.2:...	 & 1.234 & 1.167  & disc &  3	  \\
20000$+$3239 &		  &   20 01 59.52&  $+$32 47 32.90   &    G8Ia  	 & 2.645 & 0.955  & shell&  14	  \\
20056$+$1834 &QY Sge	  &   20 07 54.62&  $+$18 42 54.50   &    G0e...	 & 1.135 & 2.017  & disc &  1	  \\
22223$+$4327 &V448 Lac	  &   22 24 31.43&  $+$43 43 10.90   &    F9Ia  	 & 4.207 & 0.274  & shell&  12,14 \\
22272$+$5435 &HD 235858   &   22 29 10.37&  $+$54 51 06.35   &    G5Ia  	 & 2.838 & 1.286  & shell&  15	  \\
22327$-$1731 &HD 213985   &   22 35 27.53&  $-$17 15 26.89   &    A0III 	 & 1.353 & 1.182  & disc &  1	  \\
23304$+$6147 &		  &   23 32 44.79&  $+$62 03 49.10   &    G2Ia  	 & 2.908 & 0.432  & shell&  7,15  \\
Z02229$+$6208&		  &   02 26 41.79&  $+$62 21 22.00   &  		 & 2.546 & 1.194  & shell&  6	  \\
	     &BD$+$39 4926&   22 46 11.23&  $+$40 06 26.30   &    B8		 & 2.046 & -0.008 & disc &  1	  \\

\hline					        						   	  	  
\end{tabular}
\begin{flushleft}
$*$WISE data indicates saturation. \\
$**$Although an RV Tauri pulsator (\cite{lloydevans99,kiss07}, but not in the GCVS) we include this 
object in the reference sample because it is a well characterized disc object.\\
(1):\cite{deruyter06}; (2):\cite{gorlova12}; (3):\cite{maas07}; (4):\cite{oudmaijer95};
(5):\cite{vanwinckel96}; (6):\cite{hrivnak05}; (7):\cite{fujii01}; (8):\cite{klochkova07}; (9):\cite{sanchez06};
(10):\cite{min13}; (11):\cite{meixner04}; (12):\cite{vanwinckel00}; (13):\cite{gauba03}; (14):\cite{omont93};
(15):\cite{bujarrabal01}; (16):\cite{hrivnak00};(17):\cite{hrivnak11}; (18):\cite{woods05}; (19):\cite{wannier90}
\end{flushleft}
\label{reftable}
\end{table*}

For 60 carefully selected reference post-AGB objects, we determined
the full SEDs from the available photometric
data by using the Vizier database \citep{vizier00}. Scaled
Kurucz models \citep{kurucz03} were determined for the given stellar parameters 
(Teff, logg and [Fe/H])
obtained from the literature and given in Table~\ref{reftable}. 
The total line-of-sight reddening was determined by minimizing the
difference between the scaled reddened atmospheric model and the
photometric data with the grid-method explained in \cite{degroote11}. 
In total we have
selected 33 disc and 27 shell sources as reference sample for our study.

\subsection{Disc sources}

In recent years a lot of observational evidence has been accumulated
that SEDs which show a near IR-excess are indicative for the presence
of a stable compact dusty disc \citep[e.g.][]{deruyter06, deroo06,
  deroo07, gielen11, hillen15}. The dust grain processing both in dust
grain growth and crystallinity is very strong, which are indicators
for the longevity \citep[e.g.][]{gielen08,gielen11} and the IR spectra
of disc sources clearly stand out compared to expanding shells. Mid-
and near-IR interferometric measurements are needed to resolve the
discs and show indeed that the very compact nature of the
circumstellar environment \citep[e.g.][]{deroo06, hillen13,
  hillen14}. In two objects the Keplerian kinematics has been resolved
via CO interferometric data obtained at the Plateau de Bure
\citep{bujarrabal15} and ALMA \citep{bujarrabal13a} interferometers.
Single disc CO observations often show a narrow CO profile, again
indicative of Keplerian kinematics rather than expansion
\citep{bujarrabal13b}. Our reference sample contains 33 disc objects
with the disc sample mainly build from
\cite{deruyter06} and more recent additions (Table~\ref{reftable}). 
WISE photometry has a saturation limit for the
very bright sources.  Sources brighter than approximately 2.0, 1.5,
-3.0 and -4.0 magnitude in W1, W2, W3 and W4 respectively, are not
reliable.  Therefore, sources for which the WISE photometry suffers
from saturation are not represented in the WISE colour-colour diagram
but are included in Table~\ref{reftable} for completeness.  These
objects are HD 44179, which is the central star of the famous Red
Rectangle nebula, HR 4049, IRAS 10456$-$5712 and IRAS 09371$+$1212
which is known as the Frosty Leonis nebula.

\subsection{Shell sources}

After the AGB phase, the expansion of the remnant CSE around a single
post-AGB star is thought to produce a detached circumstellar shell. These
objects are expected to display a double-peaked SED. The expansion
velocity of the gas as well as the mass-loss rates can be determined by CO observations
\citep{bujarrabal99}. A typical expansion velocity for the circumstellar
shell is 10 to 15 km/s \citep{hrivnak00,he14}. We made an extensive
literature survey for optically bright objects for which the envelope
kinematics are studied using the rotational transition of CO 
(see Table~\ref{reftable}). We selected 27 post-AGB stars which 
show a clear double peak SED and an CO expansion velocity 
between 10 and 15 km/s and we characterise them as {\sl shell} sources.

\subsection{WISE colour-colour diagram of the reference sample}

\begin{figure}
\begin{center}
\includegraphics[bb= 0 0 1200 800, width=9cm,angle=0]{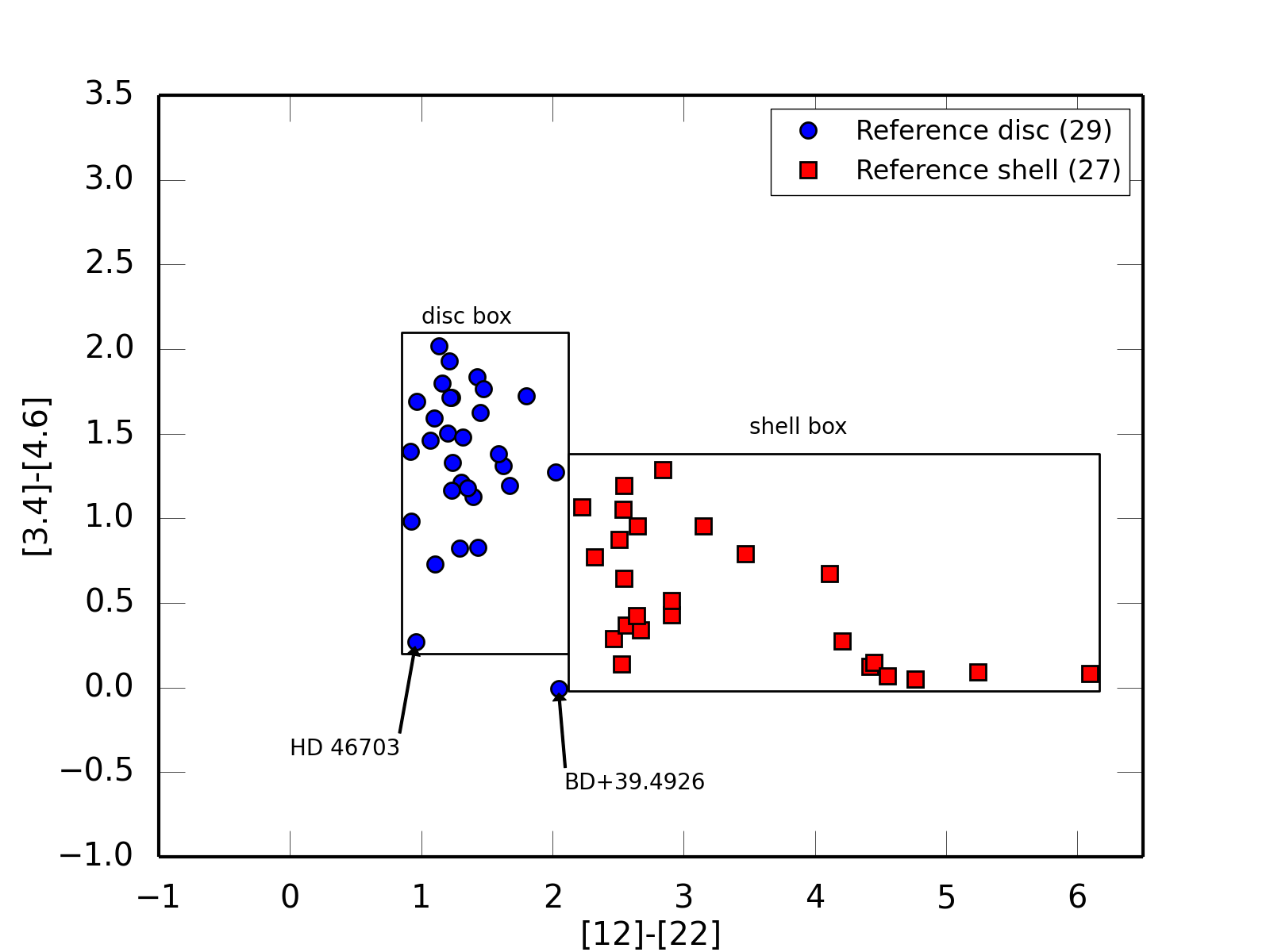}
\caption{The WISE colour-colour diagram for the 56 reference sample is depicted here.}  
\end{center}
\end{figure}

We display the WISE colour-colour diagram for our reference sample in
Fig.~3. The 29 reference disc sources and 27 reference shell sources
are represented with different symbols. The figure shows that
our reference disc sources occupy a specific region in the WISE
colour-colour diagram, which we define as the {$\it{disc}$}
{$\it{box}$ (see Fig.~3). While determining the box, HD\,46703 is
included in the box, but BD+39$^{\circ}$4926 is not.  
HD\,46703 shows several excess points in
its SED, and these are in various photometric systems (see Fig.~4). Its
excess is thus reliably detected. In addition, it is a well studied,
chemically peculiar binary object
\citep{hrivnak08}. BD+39$^{\circ}$4926, on the other hand, only
exhibits a small excess at 22~$\mu$m, only detected in W4. We will
discuss this source thoroughly in Sect.~5.

\begin{figure}
\begin{center}
\includegraphics[bb= 0 0 1200 800, width=9cm,angle=0]{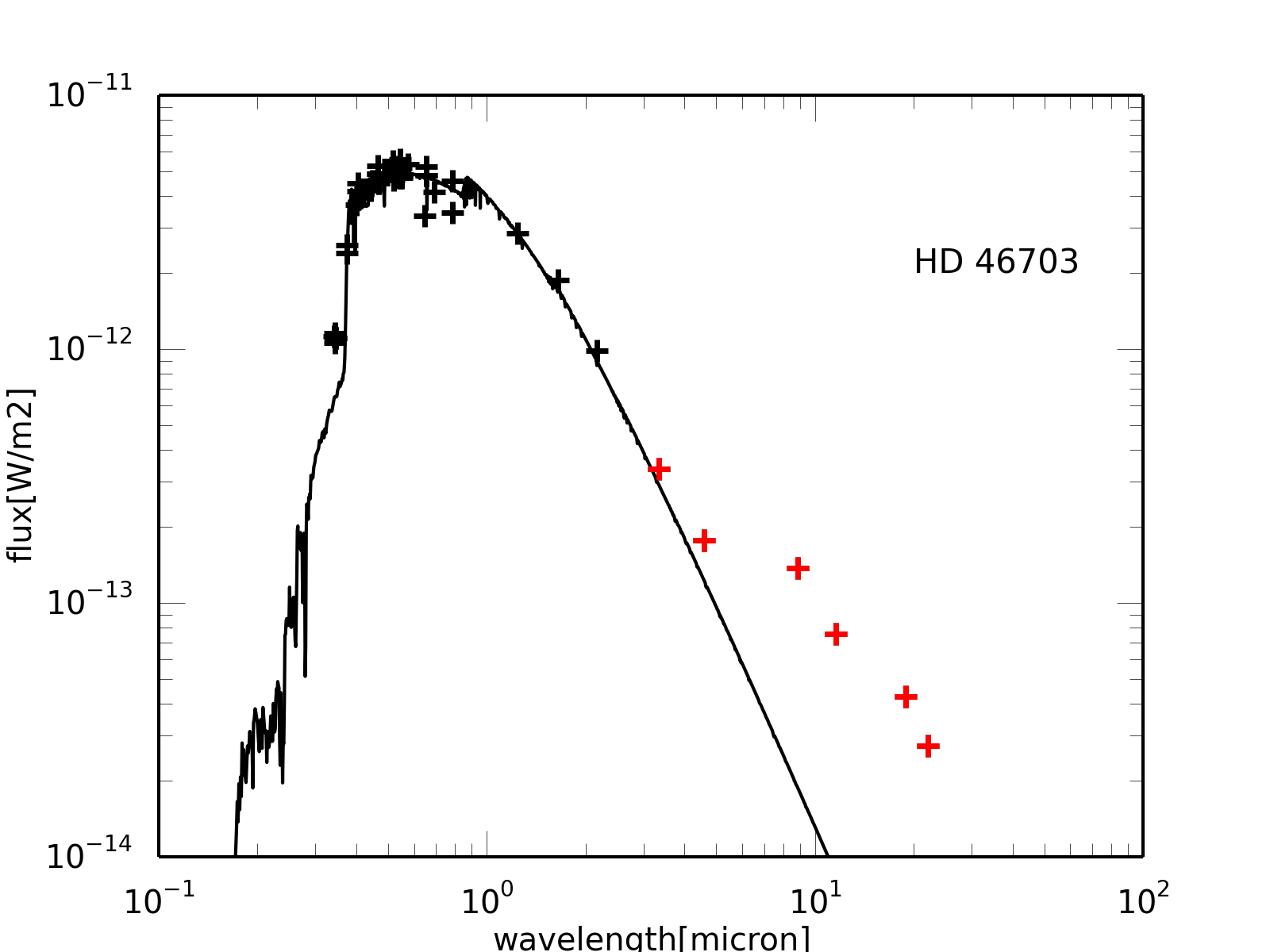}
\caption{The SED of HD 46703. The symbols are the same as in Fig. 2}  
\end{center}
\end{figure}

Reference shell sources occupy a larger
area in the diagram. It is much harder to define them with a simple
colour combination, because young shell sources may still have dust
close enough to the stars as to show a [3.4]$-$[4.6] excess. For the
evolved shells, the near-IR excess has disappeared, and they
are expected to show only a red [12]$-$[22] colour. Consequently,
we defined a big {$\it{shell}$} {$\it{box}$ which includes all
reference shell sources.

\section{RV Tauri Stars}

We compare the infrared properties of RV Tauri stars with the
reference sample in the WISE colour-colour diagram.  In Fig. 5 we
display the WISE colour-colour diagram for all 153 Galactic RV\,Tauri
stars, which are given by the GCVS.  However, 25 of the 153 RV\,Tauri
stars are probably misclassified objects \citep{IBVS91} because of
their period range and spectral types. They are not
included in our further analyses and we continue with 128
RV Tauri stars. Additionaly, for two well known RV Tauri stars, which are R
Sct and V390 Vel, the WISE photometry suffers from saturation. These two 
objects are therefore are not included in the WISE colour-colour diagram either but are included in Table~\ref{rvtauminitable}
for completeness.

\begin{table*}
 \centering
  \caption{A part of the RV Tau objects is given here. The full table
    is available online via CDS.}
  \begin{tabular}{llllcccc}
  \hline
   Name          & RA& DEC  & Spectral Type    & [12]$-$[22] & [3.4]$-$[4.6]& SED &   Ref \\
  \hline

2MASS J05271262$-$6735074   & 05 27 12.62       & $-$67 35 07.5        &              &  1.527       & 0.108     & uncertain            &                                 \\
2MASS J11092679$-$6049285   & 11 09 26.793      & $-$60 49 28.56       & 	      &  2.460	     & $-$0.089  & uncertain		&				  \\
2MASS J15183614$+$0204162   & 15 18 36.15       & $+$02 04 16.3        & F5 	      &  $-$0.131    & 0.058	 & non$-$IR excess	&				  \\
2MASS J16571174$-$0403596   & 16 57 11.74       & $-$04 03 59.7        & F5 	      &  0.298	     & 0.098	 & non$-$IR excess	&				  \\
2MASS J19163578$+$3011388   & 19 16 35.78       & $+$30 11 38.8        & G0.2:e...    &  0.542	     & 0.095	 & non$-$IR excess	&				  \\
2MASS J21333241$-$0049057   & 21 33 32.41       & $-$00 49 05.8        & F9 	      &  1.450	     & 0.079	 & uncertain		&				  \\
AA Ari  		    & 02 03 37.73846    & $+$22 52 22.2458     & K7 	      &  0.125	     & 0.185	 & non$-$IR excess	&				  \\
AC Her  		    & 18 30 16.23850    & $+$21 52 00.6080     & F2Iep        &  1.271	     & 1.448	 & disc 		&  1	                           \\
AD Aql  		    & 18 59 08.696      & $-$08 10 14.12       & G8 	      &  1.694	     & 1.437	 & disc 		&  1	                           \\
AI Sco  		    & 17 56 18.5733     & $-$33 48 43.359      & G5 	      &  0.923	     & 1.610	 & disc 		&  1	                           \\

\vdots & \vdots & \vdots & \vdots & \vdots & \vdots & \vdots & \vdots \\
\hline					        						   	  	  
\end{tabular}
\begin{flushleft}
(1):\cite{deruyter06}
\end{flushleft}
\label{rvtauminitable}
\end{table*}

RV Tauri stars cluster in three
different regions in the WISE colour-colour diagram (see Fig.~5). In
order to check this clustering we obtained full SEDs of
all RV\,Tauri stars using the Vizier database as a source for the 
photometric data \citep{vizier00}. For some of them,
we used the appropriate scaled Kurucz model \citep{kurucz03} 
with the stellar parameters taken from the literature (see the references in Table~\ref{rvtauminitable}).
For the sources where we lack indicative stellar parameters, we only analysed the reddened data.  
We note that the SEDs are affected by the large-amplitude pulsations.

\begin{figure}
\begin{center}
\includegraphics[bb= 0 0 1200 800, width=9cm,angle=0]{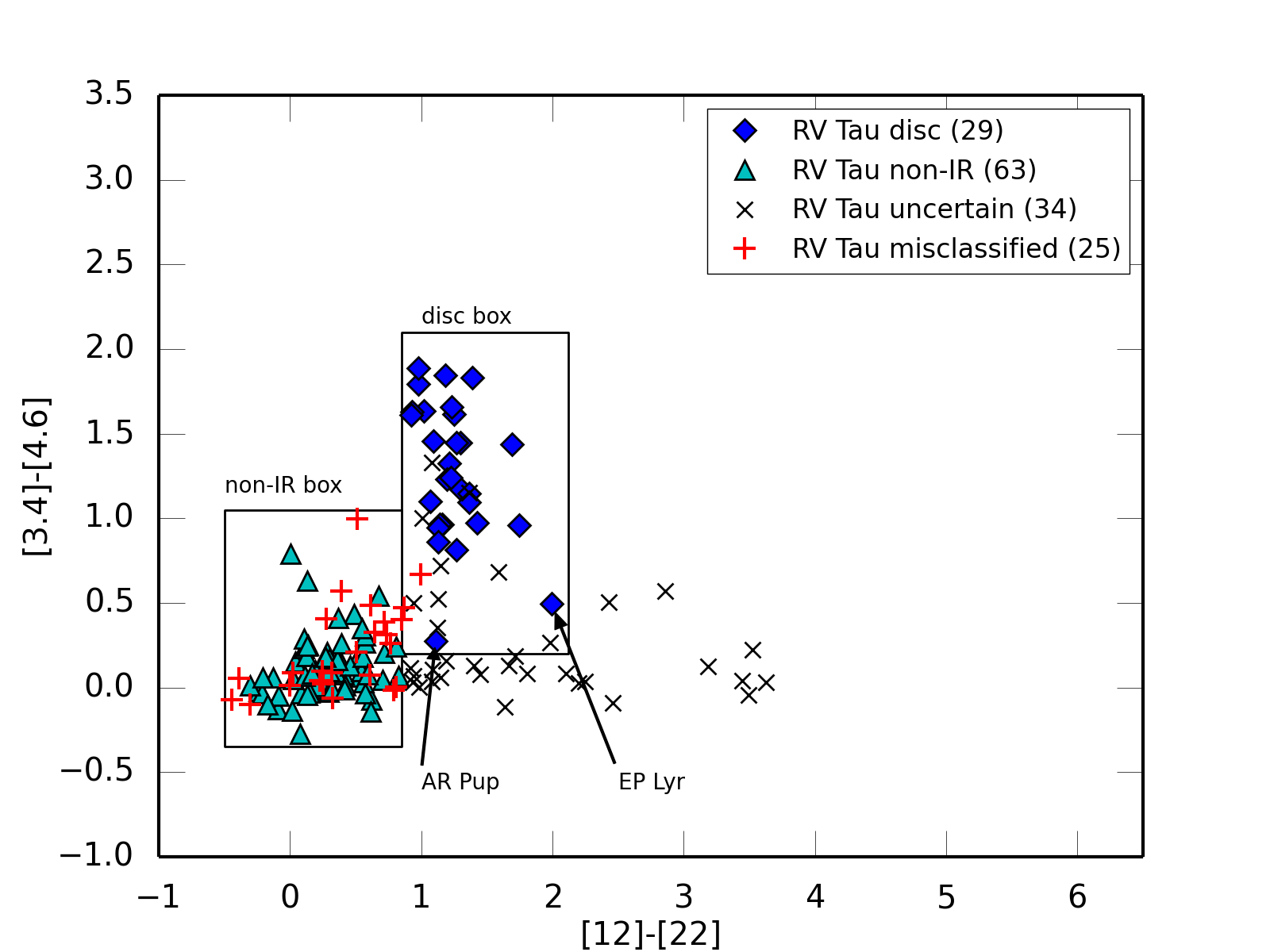}
\caption{Three different type of SED characteristic among the Galactic RV Tauri stars and probably misclassified RV Tauri stars are represented here with different symbols.}  
\end{center}
\end{figure}

The first striking clump is seen in the previously defined disc
box. This box contains 38 RV Tauri, 29 of which show a clear disc
SED. These are shown with diamonds in Fig.5.  This clump is the WISE
alternative of the RV Tauri box as defined by \cite{evans85}.  For the
remaining 9 stars in the box, the full SED is not clear, which is
likely due to the large amplitude of the pulsations in combination
with a poor photometric sampling, which makes that the SED is not well
defined. Several of these sources may actually have a disc excess, but
pulsation and the poor wavelength sampling makes their SEDs become
unclear. Therefore, we classify these sources as {$\it{uncertain}$}.

Another remarkable clumping can be seen around the zero point
of the diagram. These RV Tauri stars constitute a big part of the 
Galactic RV\,Tauri population. In total 63 RV Tauri stars occupy this 
region which we call the  $\it{non-IR}$ $\it{box}$.

Finally, some of the RV Tauri stars occupy the region which we
defined as  the shell box in the previous section. They show a mild [3.4]-[4.6] 
excess with 2.2 $\leq$ [12]-[22] $\leq$ 3.7, which indicates the presence of cool dust, characteristic
of shell sources.  However, they do not show a clear double-peaked SED
as do the reference shell sources: in most cases only the [22] micron
band is in excess. They could be potential shell sources but, due to
the absence of information about envelope kinematics of these stars
and the lack of more data in the IR to confirm the dusty excess, we
prefer to classify them as {$\it{uncertain}$} sources.

After carefully examining the full SEDs we conclude that indeed all 29 (23${\%}$) RV
Tauri stars which cluster in the disc box display a broad IR-excess in their
SEDs. Many of them have already been given a disc status in the literature
\citep{deruyter06,deruyter05}. The 63 (50${\%}$) RV Tauri
stars, in the non-IR box do not display any IR excess at all, also not when considering their full SED. 
We called them as {$\it{non-IR}$} sources. 
Finally, 34 (27${\%}$) RV Tauri stars have an unclear SED and 
we label them as {$\it{uncertain}$} sources.
The three groups are represented with different symbols in Fig. 5.

\section{ANALYSIS}

In our analyses we extensively used the literature to correlate
SED characteristics with other observables. We focus on binarity
as well as the presence of photospsheric chemical anomalies. We
use the reference sample to investigate the Galactic RV Tauri population.

\subsection{Disc and Binarity}

\begin{table}
 \centering
  \scriptsize
  \caption{Confirmed binaries among our whole sample. We give in the second column
    the SED types discussed in the main text.}
  \begin{tabular}{lccc}
\hline
Name            & Type     & $P_{\rm orbital}$(day)        &  Ref  \\	
\hline
89 Her		  &    pagb-disc & 	288.4  &     12  \\
BD$+$39 4926	  &    pagb      & 	775.0  &     6   \\
BD$+$46 442	  &    pagb-disc  & 	140.8  &     4   \\
EN TrA		  &    pagb-disc & 	1490.0 &     13   \\
HD 108015	  &    pagb-disc & 	938.0  &     5   \\
HD 213985	  &    pagb-disc & 	259.0  &     1   \\
HD 44179	  &    pagb-disc & 	298.0  &     9   \\
HD 46703	  &    pagb-disc & 	610.0  &     5   \\
HD 52961	  &    pagb-disc & 	1310.0 &     11  \\
HD 95767	  &    pagb-disc & 	2050.0 &     5   \\
HR 4049 	  &    pagb-disc & 	429.0  &     9   \\
IRAS05208$-$2035  &    pagb-disc & 	236.0  &     2   \\
IRAS15469$-$5311  &    pagb-disc & 	389.9  &     13  \\
IRAS17038$-$4815  &    pagb-disc &      1381.0 &     1   \\
IRAS19157$-$0247  &    pagb-disc & 	120.5  &     13  \\
SAO 173329	  &    pagb-disc & 	115.9  &     1   \\
ST Pup		  &    pagb-disc & 	410.0  &     3   \\
AC Her		  &    RVTauri-disc & 	1196.0 &     10  \\
BD$+$03 3950	  &    RVTauri-disc & 	519.6  &     13   \\
IRAS09144$-$4933  &    RVTauri-disc & 	1770.0 &     1   \\
RU Cen		  &    RVTauri-disc & 	1489.0 &     1   \\
SX Cen		  &    RVTauri-disc & 	600.0  &     1   \\
U Mon		  &    RVTauri-disc & 	2597.0 &     8   \\
V390 Vel	  &    RVTauri-disc & 	499.0  &     7   \\

\hline					        						   	  	  
\end{tabular}
\begin{flushleft}
(1):\cite{gielen08}; (2):\cite{gielen11}; (3):\cite{gonzalez96}; (4):\cite{gorlova12};
(5):\cite{vanwinckel00b}; (6):\cite{kodaira70}; (7):\cite{maas03}; (8):\cite{pollard95}; (9):\cite{vanwinckel95};
(10):\cite{vanwinckel98}; (11):\cite{waelkens92};
(12):\cite{waters93}; (13):\cite{vanwinckel09}

\end{flushleft}
\label{binarytable}
\end{table}

\begin{figure}
\begin{center}
\includegraphics[bb= 0 0 1200 800, width=9cm,angle=0]{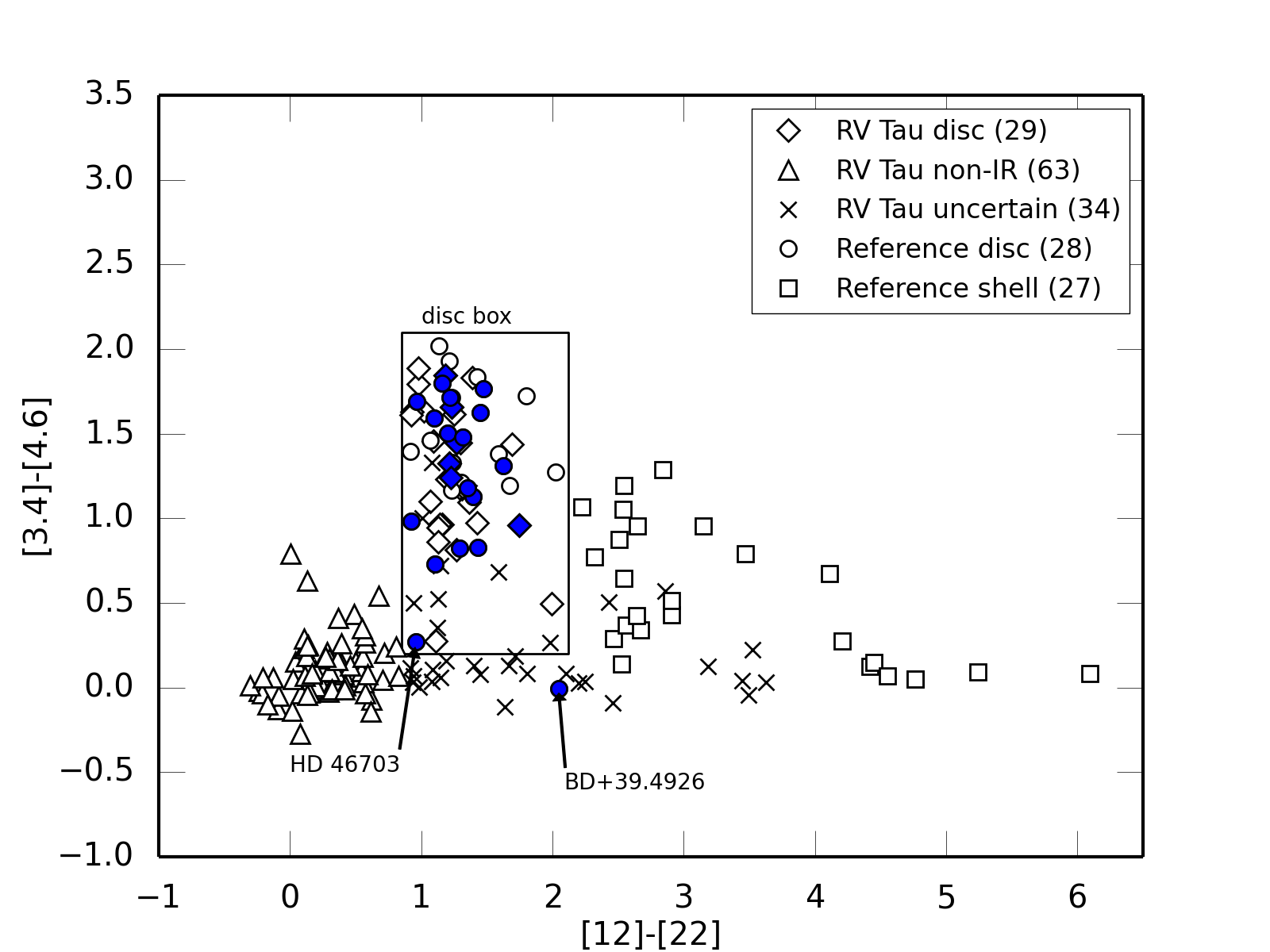}
\caption{WISE [12]$-$[22], [3.4]$-$[4.6] colour-colour diagram of the 
merged reference$+$RV Tauri sample. Confirmed binares are shown in blue.}
\end{center}
\end{figure}

There are 17 confirmed binaries among our reference sample. We
indicated these 17 objects with blue circles in the WISE
colour-colour diagram in Fig.~6. For the reference stars, the
correlation between binarity and the presence of a disc SED is
generally acknowledged \citep{vanwinckel09}. Fig. 6 illustrates
this statement most clearly, as the binary post-AGB objects are 
clearly clustered in the disc region of the diagram.

Direct detection of orbital motion in RV Tauri stars by means of
radial velocity measurements is difficult because of the significant
pulsational amplitude. Despite this difficulty, orbital elements have
been determined for 7 RV Tauri stars so far. In Fig.~6 these binary RV
Tauri stars are also represented, but with blue diamonds. The same
binarity$-$disc SED correlation can also be seen for the RV Tauri
stars. All confirmed binaries are listed in Table~\ref{binarytable}

RV Tauri photometric class b objects show a flux variation with a period much
longer than their pulsation period.  \cite{vanwinckel99} has
explained the RVb phenomenon by assuming that a circumbinary dusty
disc is present and that the viewing angle onto the disc determines 
the photometric class. If the disc is seen under a high inclination, 
long-period variation may be seen due to variable circumstellar 
extinction during the orbital motion. Under this interpretation, the long-term period 
in the light curve is then the orbital period. If the system is seen with a small
inclination (near pole-on), the line-of-sight is not obscured and no 
long-period variations will be detected. 
We have searched for RVa and RVb type classifications in the
literature \citep{kiss07,rao14}. We found 11 RVb and 20 RVa type objects. In Fig.~7  the RVa and
RVb type RV Tauri stars are indicated with blue and red squares, respectively, 
in the WISE colour-colour diagram (R Sct is not included in the Fig.~7, since its WISE photometry suffers from saturation).
All RVb stars are in the disc box and none
are outside it. This corroborates the finding that the RVb phenomenon is 
related to the presence of circumstellar material. 
Discs with the correct inclination appear
as RVb in the disc box. For 2 of 11 RVb objects (U Mon and SX Cen) and
for 2 of 20 RVa objects (AC Her and RU Cen) the binary nature has been
proven so far. An important  difference between RVa and RVb type 
objects may very well be our viewing angle onto their circumstellar dust. More radial velocity
monitoring is needed for all RVa and RVb objects to test their binary
nature.

\begin{figure}
\begin{center}
\includegraphics[bb= 0 0 1200 800, width=9cm,angle=0]{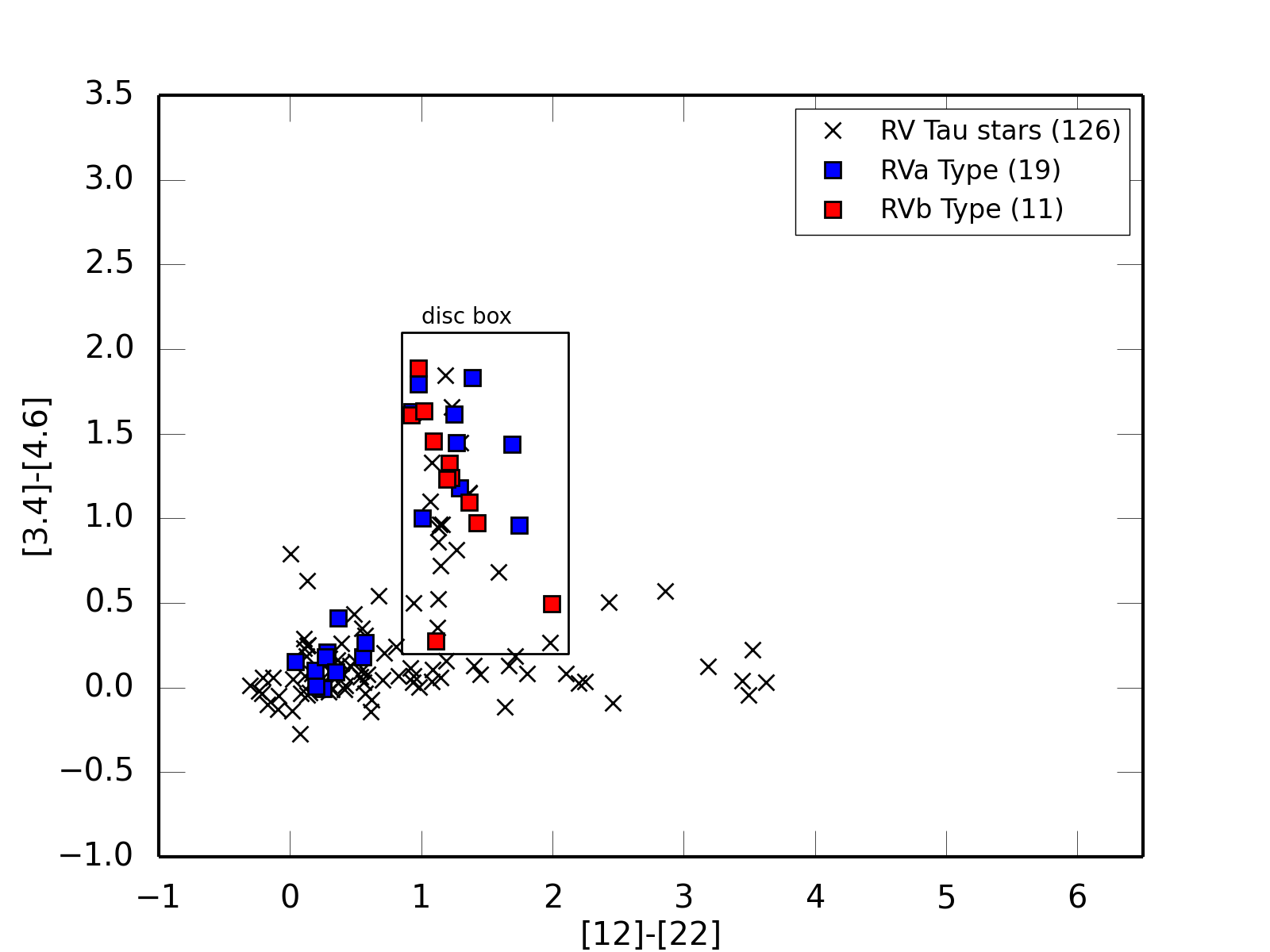}
\caption{Photometric classification is represented with different colours in the WISE colour-colour diagram.}  
\end{center}
\end{figure}

\subsection{Disc and Chemical Depletion}

In this study, one of our goals is to investigate whether there is
a relation between infrared properties and the presence of the 
photospheric chemical anomaly called depletion.
The most characteristic chemical signatures of depleted photospheres are high
[Zn/Fe], [Zn/Ti] and [S/Ti] ratios. \cite{waters92} proposed that
the most likely circumstance allowing the process occurs when the circumstellar 
dust is trapped in a circumstellar disc. Very strongly depleted atmospheres 
were first found  exclusively in binary post-AGB objects
\citep{vanwinckel95}, but was found to be
widespread in the last decennia.}

As the [Zn/Ti] and [S/Ti] ratio's are good tracers
for depleted photospheres, we used these abundance ratios, depending 
on their availability in the literature, as a tracer of depletion in our sources.
We defined the objects as follows: if
objects have [Zn/Ti] or [S/Ti] ratios larger than 1.5, they are
classified as {$\it{strongly}$} {$\it{depleted}$}. If their [Zn/Ti]
or [S/Ti] ratio is in the range 1.5$-$0.5, we classified them as
{$\it{mildly}$} {$\it{depleted}$}. Finally, if objects have [Zn/Ti]
or [S/Ti] ratio smaller than 0.5, we defined that there is no
evidence for photospheric depletion and we labelled them as {$\it{not}$}
{$\it{depleted}$}. The depletion quantifications are listed
in the Table~\ref{depletiontable}.

\cite{waters92} proposed that the most probable
location for the depletion process to take place is in a dusty stable
disc. For post-AGB stars, a stable disc is probably only present if
the star resides in a binary system, as a orbiting disc can only be
formed when the evolved star interacts closely with its companion. In a
single post-AGB star, the mass loss in the AGB wind is expected to expand.
Photospheric depletion patterns are observed in our reference
post-AGB sample as in RV\,Tauri stars.

In Fig.~8 we present the depletion quantifications with different
colours only for the reference sample. Strongly and mildly depleted
objects are mostly concentrated in the disc region of the
diagram. However, not all disc sources show depletion, as several disc
objects are classified as not depleted.  Moreover, there are 17
confirmed binary objects in the disc box of the WISE colour-colour
diagram. All these binaries show a clear broad band SED (except
BD+39$^{\circ}$4926, which shows a small excess only in the WISE W4
band, see section 5) but for 4 of them there is no evidence for
depletion. The only clear observable from Fig.~8 is that among the
shell sources depletion is absent, so far and that the binarity is likely a 
necessary but not a sufficient requirement for depletion to occur.

In Fig.~9 we illustrate the depletion quantification for RV Tauri
stars only. We show that RV Tauri stars follow the same trend as the
reference post-AGB objects. Strongly and mildly depleted RV Tauri
stars are mainly clustered in the disc box of the WISE colour-colour
diagram. There are, however, also not-depleted objects in
the disc box. One of these not-depleted objects is U Mon, which is
a confirmed binary. Another interesting outlier is SS
Gem. It is a strongly depleted object but it does not display a clear IR
excess. All 7 RV Tauri stars whose orbital parameters have been measured so
far show a clear disc SED, but not all of them are depleted
and there are depleted objects without a dust excess. 

\begin{figure}
\begin{center}
\includegraphics[bb= 0 0 1200 850, width=9cm,angle=0]{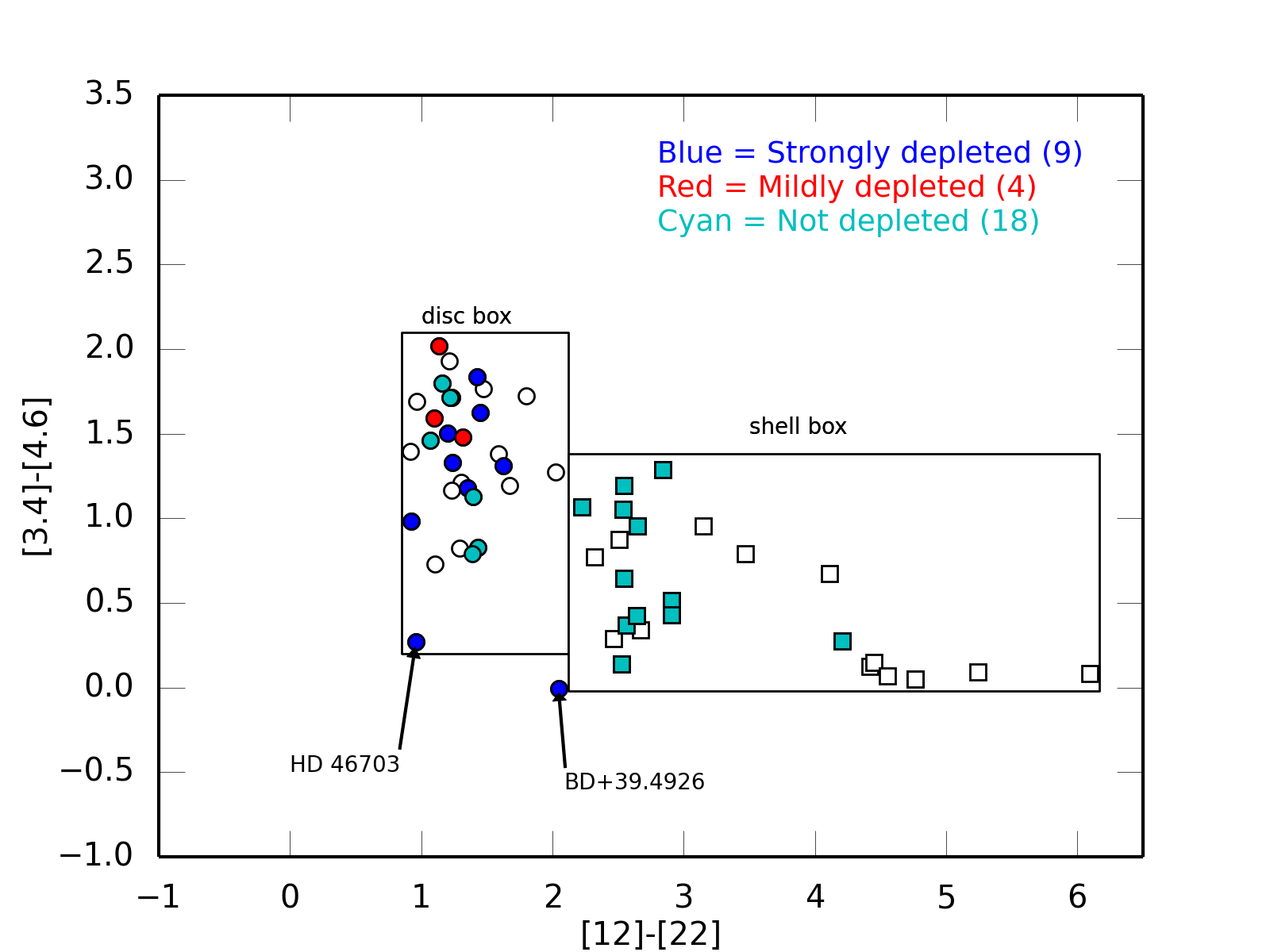}
\caption{Depletion quantifications are shown with different colours for the reference sample.}  
\end{center}
\end{figure}

\begin{figure}
\begin{center}
\includegraphics[bb= 0 0 1200 800, width=9cm,angle=0]{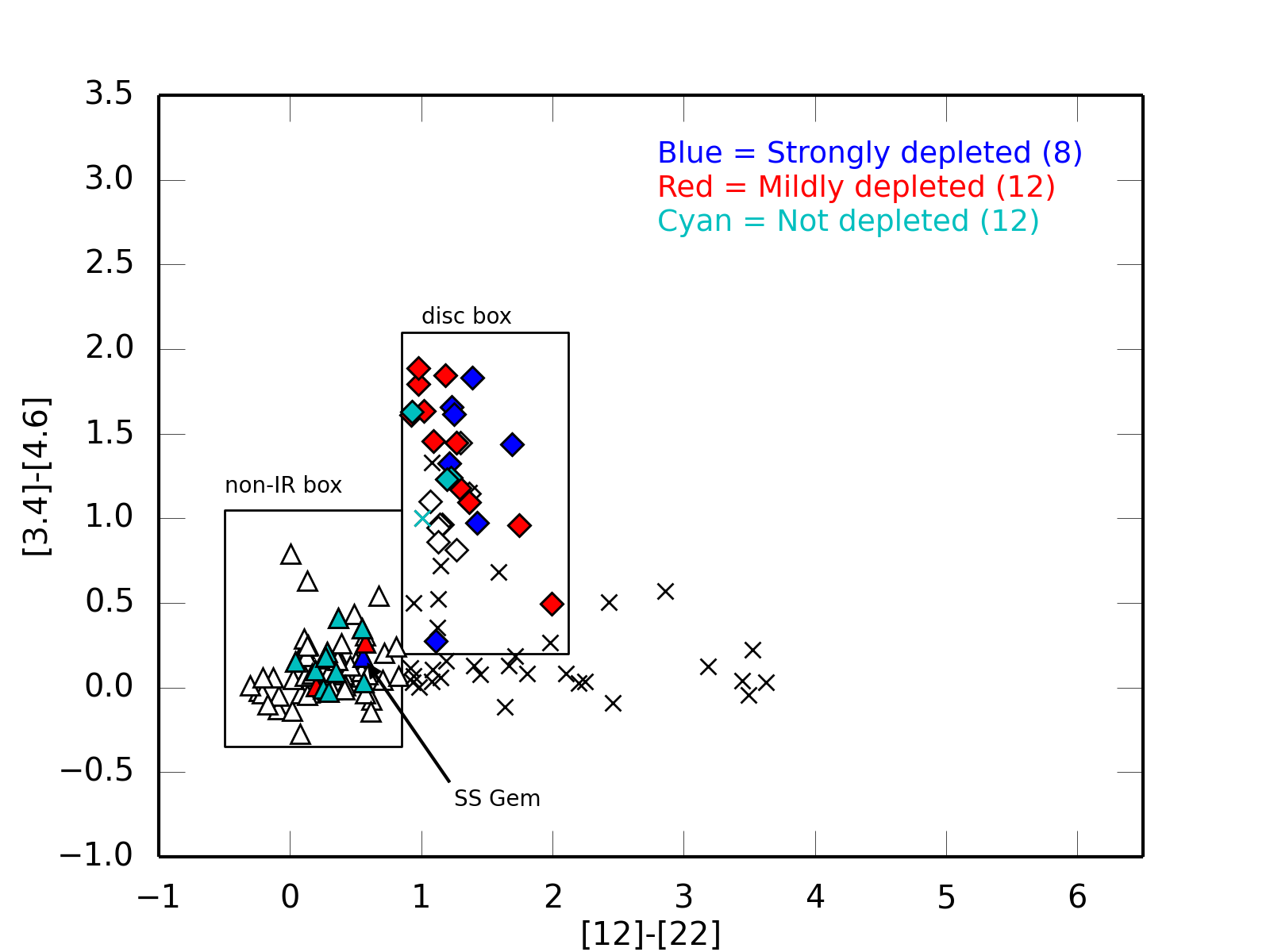}
\caption{Depletion quantifications are shown with different colours for the Galactic RV Tauri stars.}  
\end{center}
\end{figure}

\begin{figure}
\begin{center}
\includegraphics[bb= 0 0 1200 800, width=9cm,angle=0]{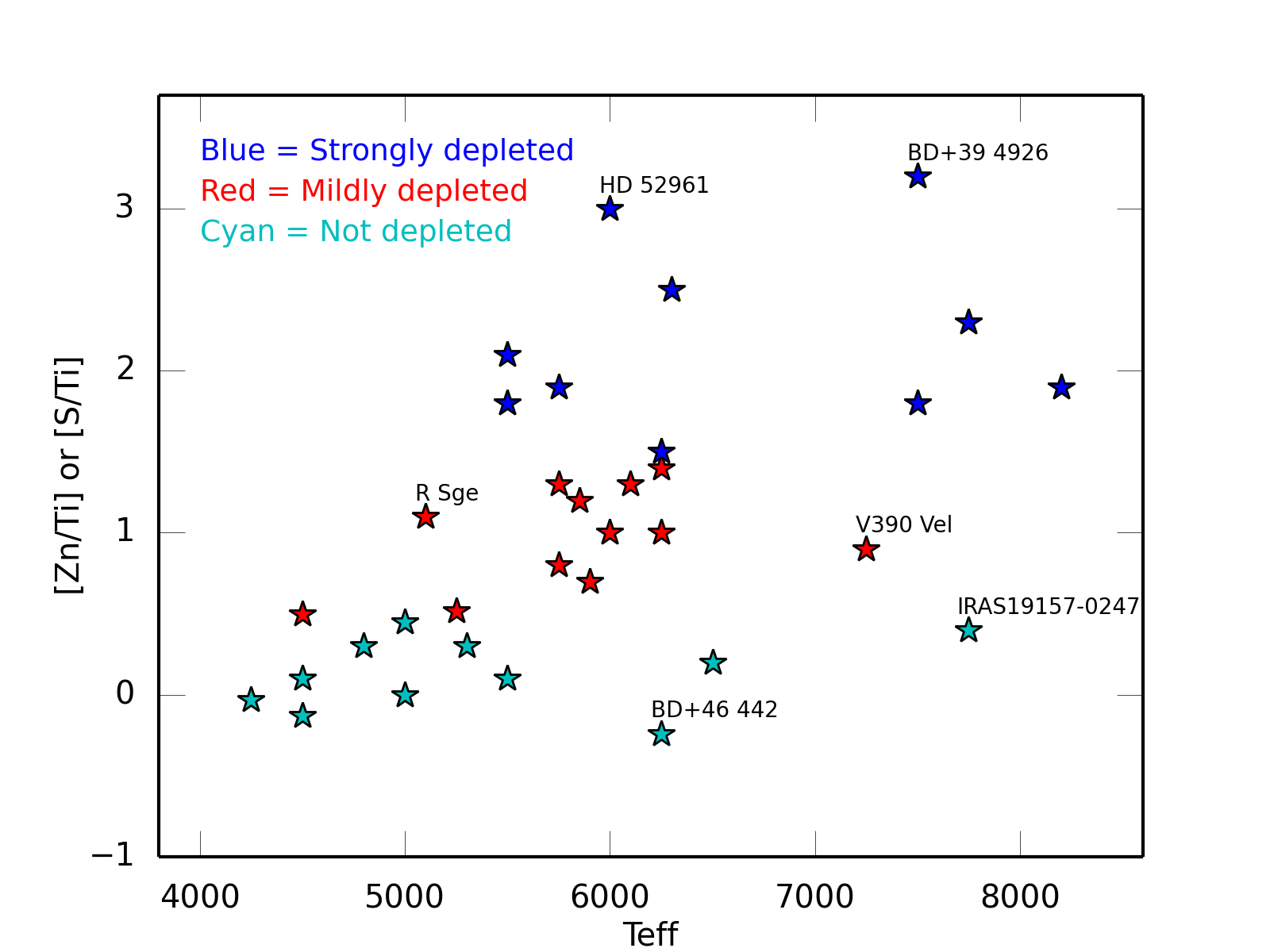}
\caption{Effective temperature versus depletion quantifications is given here.}  
\end{center}
\end{figure}

\begin{table}
 \centering
 \scriptsize  
  \caption{Detailed information of the RV Tauri stars which 
are chemically studied up to now. $[F/H]_{\rm 0}$ gives the estimated
initial metallicity obtained via the Zn or S abundance, 
PC gives the photometric class and SC gives the spectroscopic class of the RV\,Tauri sample star.}
  \begin{tabular}{lrcccccl}

\hline
  Name           & $[F/H]_{\rm 0}$   & PC  & SC    & $T_{\rm eff}$ & Binarity & SED &  Ref \\
\hline
   \multicolumn{7}{c}{Strongly depleted RV\,Tauri Stars.} \\
 \hline
 
AD Aql		    	    &	$-$0.10	  &   a &  B  &   6300 &	   & disc      & 	2 \\		
AR Pup		    	    &	0.40 	  &   b &  B  &   6000 &	   & disc      & 	5b,8 \\
BD$+$03 3950	    	    &	0.10 	  &   ? &  ?  &   7750 & Binary	   & disc      & 	6 \\
CT Ori		            &	$-$0.60   &   a &  B  &   5500 &           & disc      & 	5a,8  \\
DY Ori		    	    &	0.20 	  &   a &  B  &   6000 &	   & disc      & 	5b,7 \\
IW Car		    	    &	0.20 	  &   b &  B  &   6700 &	   & disc      & 	1,8 \\
SS Gem		  	    &	$-$0.20	  &   a &  A  &   5600 &	   & non$-$IR  & 	5a \\
SX Cen		    	    &	$-$0.30	  &   b &  B  &   6250 & Binary    & disc      & 	6 \\

\hline
 \multicolumn{7}{c}{Mildly depleted RV\,Tauri Stars.} \\
 \hline 

AC Her		    	    &  $-$0.90  &   a &  B  &   5900 & Binary	 & disc      &   2  \\
AZ Sgr		  	    &   	&   a &  A  &   4750 &	         & non$-$IR  &	4,7\\
BT Lac		    	    &  $-$0.10  &   b &  A  &   5000 &		 & disc      &   7  \\
EP Lyr		    	    &  $-$0.90  &   b &  B  &   6100 &		 & disc      &   5b,7  \\
EQ Cas		  	    &  $-$0.30  &   a &  B  &   4500 &		 & non$-$IR  &   4  \\
IRAS09144$-$4933    	    &  0.00     &   ? &	 ?  &   5750 & Binary	 & disc      &  6   \\
R Sge		    	    &  0.10     &   b &  A  &   5100 &		 & disc      &   5b  \\
RU Cen		    	    &  $-$1.10  &   a &  B  &   6000 & Binary	 & disc      &   6  \\
RV Tau		    	    &  $-$0.40  &   b &  A  &   4500 &		 & disc      &   3  \\
SU Gem		    	    &  0.00     &   b &  A  &   5250 &		 & disc      &  7   \\
UY Ara		    	    &  $-$0.30  &   a &  B  &   5500 &		 & disc      &   3,7  \\
V390 Vel	    	    &  0.00     &   ? &	 ?  &   7250 & Binary	 & disc      &   6  \\

\hline
 \multicolumn{7}{c}{Not depleted RV\,Tauri Stars.} \\
\hline

AI Sco		  	 &   $-$0.70     &   b &  A  &   5300 &	      & disc	   &	4,8\\
AR Sgr		  	 &   	         &   a &  A  &   5300 &	      & non$-$IR   &	4,7\\
BT Lib		  	 &   $-$1.10     &   ? &  C  &   5800 &	      & non$-$IR   &	3\\
DF Cyg                   &               &   a &  A  &   4800 &	      & uncertain  &	4\\
DS Aqr                   &               &   a &  C  &   6500 &	      & non$-$IR   &	3\\
GP Cha 	          	 &   $-$0.60     &   b &  A  &   5500 &	      & disc	   &	6,7\\
R Sct		  	 &   $-$0.20     &   a &  A  &   4500 &	      & uncertain  &	3\\
RX Cap		  	 &   $-$0.60     &   ? &  A  &   5800 &	      & non$-$IR   &	4\\
TT Oph		  	 &   $-$0.80     &   a &  A  &   4800 &	      & non$-$IR   &	3\\
TW Cam		         &   $-$0.40     &   a &  A  &   4800 &       &	disc       &	 3\\
TX Oph		  	 &   $-$1.10     &   ? &  A  &   5000 &	      & non$-$IR   &	4\\
U Mon		  	 &   $-$0.50     &   b &  A  &   5000 & Binary& disc	   &	3,8\\
UZ Oph		  	 &   $-$0.80     &   a &  A  &   5000 &	      & non$-$IR   &	4\\
V Vul		  	 &   0.10        &   a &  A  &   4500 &	      & disc	   &	4,7\\
V360 Cyg	  	 &   $-$1.30     &   a &  C  &   5300 &	      & non$-$IR   &	2\\
V453 Oph	  	 &   $-$2.20     &   a &  C  &   5800 &	      & non$-$IR   &	2\\
V820 Cen	  	 &   $-$2.20     &   a &  C  &   4750 &	      & non$-$IR   &	7\\

\hline	

\end{tabular}
\begin{flushleft}
(1):\cite{giridhar94}; (2):\cite{giridhar98}; (3):\cite{giridhar00}; (4):\cite{giridhar05};
(5a):\cite{gonzalez97a}; (5b):\cite{gonzalez97b}; (6):\cite{maas05}; (7):\cite{rao14}; 
(8):\cite{kiss07}
\end{flushleft}
\label{rvdepletiontable}
\end{table}

\cite{giridhar05} have suggested that depletion is mostly observed in
stars hotter than 5000 K. As the atmospheres of cooler stars have a
deep convective envelope that mixes the accreted material with the whole
convective part of the envelope and hence dilutes the effect of
depletion. \cite{venn14} has also suggested the same scenario to
explain not depleted RV Tauri stars with a disc. One can expect a
lower depletion trend in function of effective temperature but it also
depends on the mass ratio between the accreted material and the
convective envelope. In Fig.~10 we illustrate effective temperature
versus depletion quantification for the disc sources. We can see a
general trend that hotter objects are, on general, somewhat more
depleted, but there is clearly no strong relation between effective
temperature and depletion.

\cite{giridhar05} has also suggested that depletion characteristics of
RV Tauri stars may depend on their initial metallicity. Field RV Tauri 
classes of both spectroscopic classes A and B show depletion abundance patterns
(see Table~\ref{depletiontable}).
The initial metallicity, $[Fe/H]_{\rm 0}$, covered by these studies ranges from $-$0.1 to $-$0.8. 
The C-type RV Tauri stars in the field and in globular
clusters do not show depletion patterns and their observed initial
metallicities are smaller than $-$1.0. It may be
indicating that the depletion process is not efficient at these low
initial metallicities \citep{giridhar00}. The initial
metallicity is, however, hard to determine in depleted objects. In 
Table~\ref{rvdepletiontable} we list the chemically studied RV Tauri stars with their
spectroscopic and photometric classes, effective temperature and
initial metallicity. The C-type RV Tauri stars
which are intrinsically metal poor objects are indeed not depleted while all
the strongly depleted and mildly depleted objects are of spectroscopic
class A or B. There are also many A-type objects
which are not depleted.

\begin{figure}
\begin{center}
\begin{tabular}{c}
\resizebox{\hsize}{!}{\includegraphics[bb= 0 0 1200 800, width=8cm,angle=0]{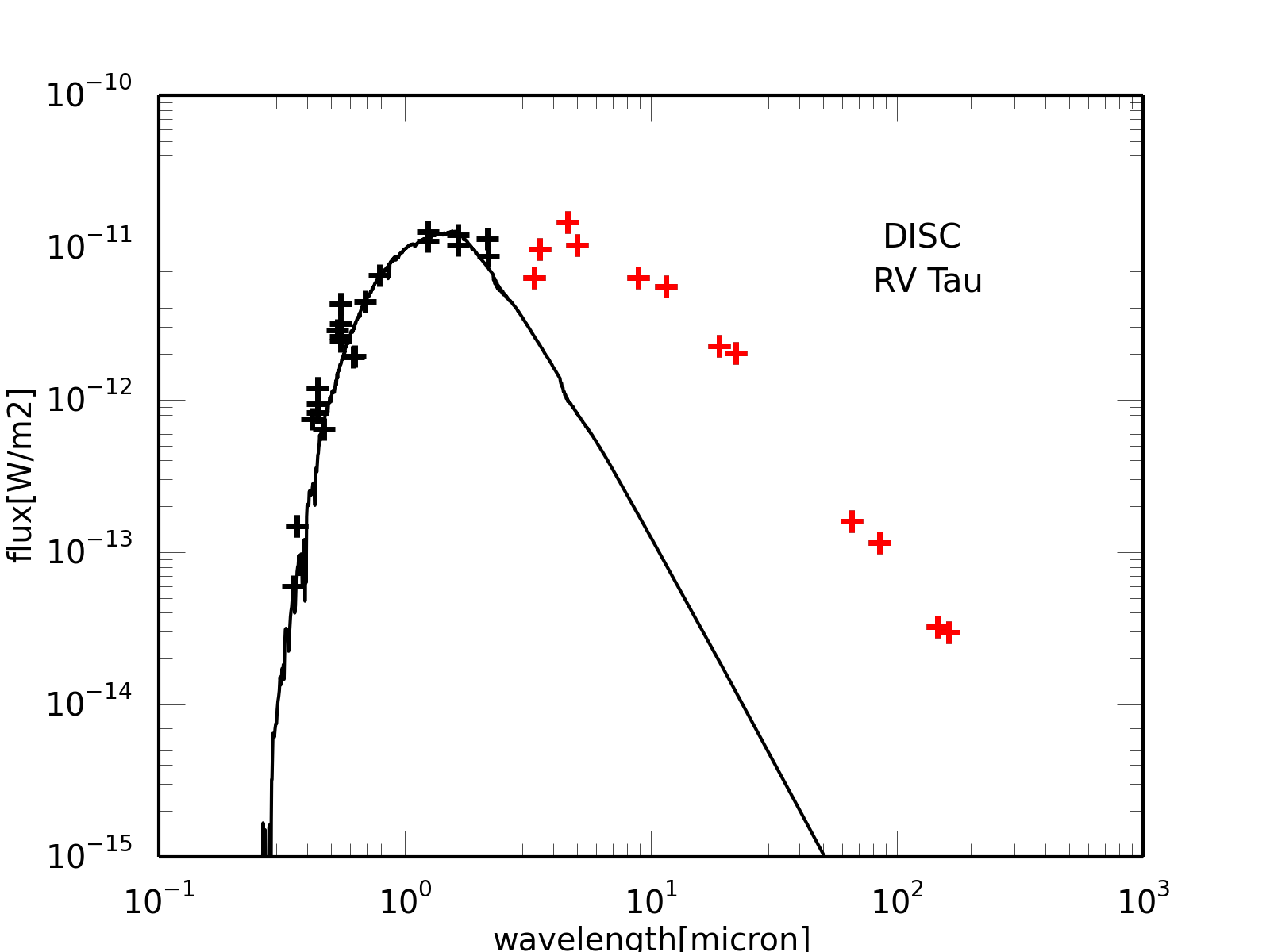}}\\\\
\resizebox{\hsize}{!}{\includegraphics[bb= 0 0 1200 800, width=8cm,angle=0]{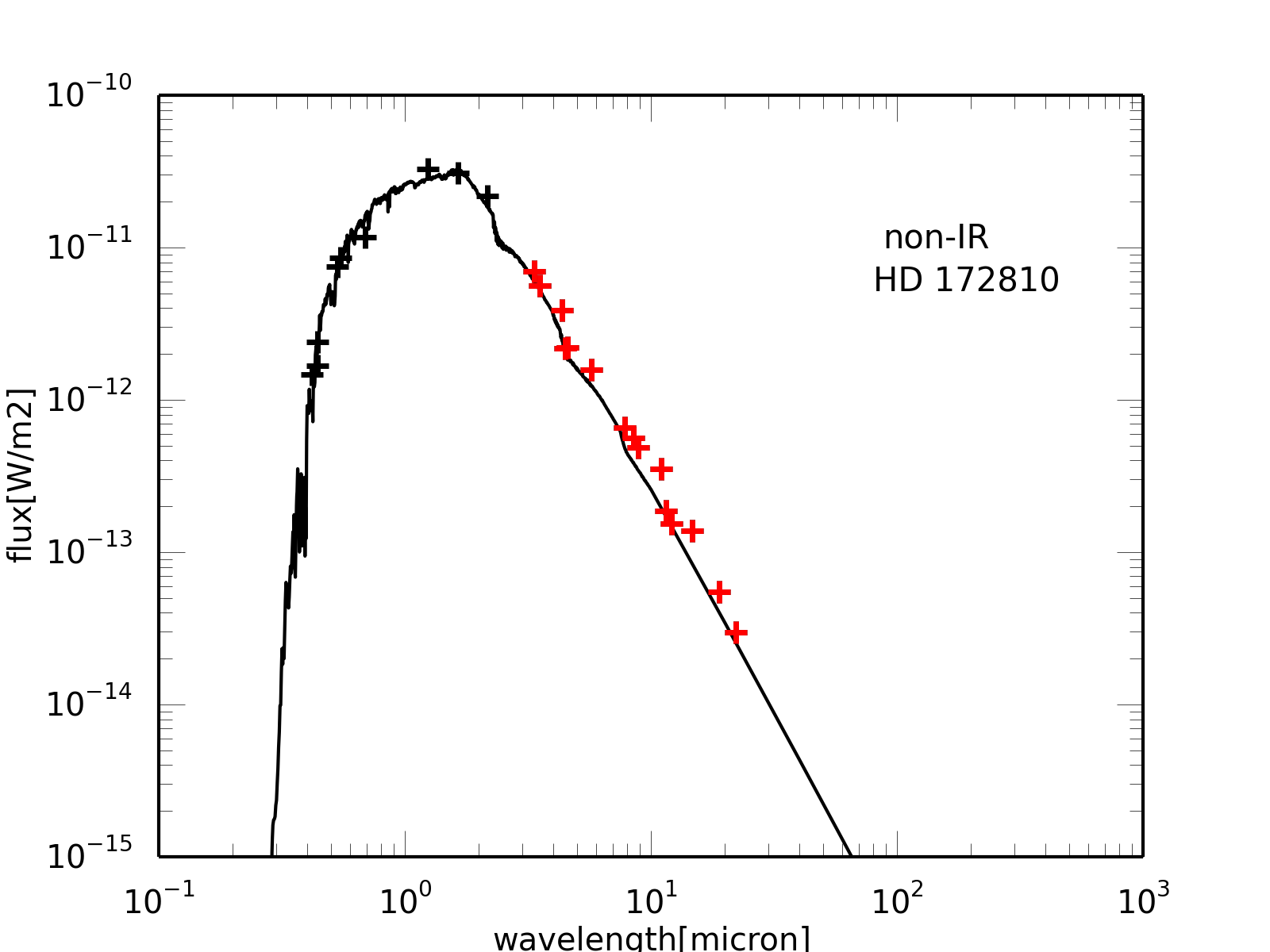}}\\\\
\resizebox{\hsize}{!}{\includegraphics[bb= 0 0 1200 800, width=8cm,angle=0]{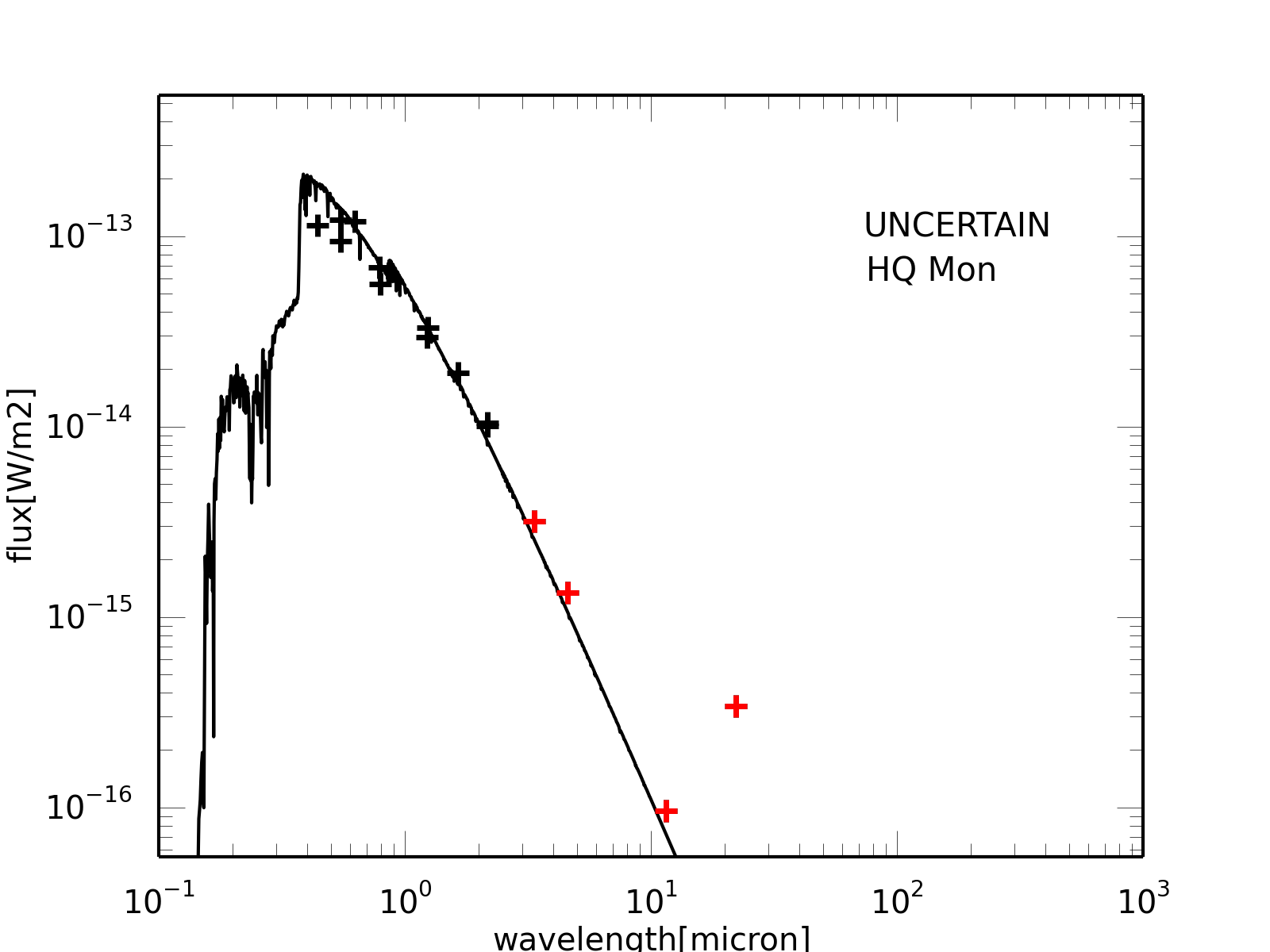}}\\
\end{tabular}
\caption{Prototypical SEDs of the three SED classes defined in this paper, in the RV Tauri case.
Top panel: RV Tau, the prototype of the RV Tau class and a clear disc object. Middle panel: HD172810, 
an example of a non-IR excess source. Bottom panel: HQ Mon, one of the uncertain sources.}
\end{center}
\end{figure}

\begin{table*}
 \centering
 \scriptsize
  \caption{Quantification of the depletion patterns obtained from the literature.}
  \begin{tabular}{lcrrrrrrrccc}
\hline

  Name             & Type     & $[F/H]_{\rm 0}$      &    [Fe/H]       &   [Zn/Ti]  &   [S/Ti]    &  [Zn/Fe] & [C/Fe] &   [s/Fe]  & SED & Binarity & Ref  \\
	
\hline
  \multicolumn{12}{c}{Strongly Depleted  Objects.} \\
\hline

 AD Aql		    &	 RV*     &   $-$0.10  &	  $-$2.10   &    2.50	 &  2.60     &   2.0   & ...  &  ...  &   disc    &  	  & 2	 \\
 AR Pup		    &	 RV*     &   0.40     &	  $-$0.90   &		 &	     &         & ...  &  ...  &   disc    &  	  & 5b	 \\ 
 BD$+$03 3950	    &	 RV*     &   0.10     &	  $-$0.30   &    2.30	 &  2.60     &   0.4   & ...  &  ...  &   disc    & Binary& 12	 \\
 BD$+$39 4926	    &	 pA*     &   $-$0.10  &	  $-$2.90   &		 &  3.20      &        & ...  &  ...  &   disc    & Binary& 10	 \\
 CT Ori		    &	 RV*     &   $-$0.60  &	  $-$1.90   &    1.90	 &  2.00      &  1.3   & ...  &  ...  &   disc    &       & 5a,22 \\
 DY Ori		    &	 RV*     &   0.20     &	  $-$2.30   &	2.10	 &  2.50      &  2.1   & ...  &  ...  &   disc    &  	  & 5b	 \\
 HD 213985	    &	 pA*     &   	      &	  $-$0.90   &		 &  1.90      &        & ...  &  ...  &   disc    & Binary& 17	 \\
 HD 44179	    &	 pA*     &   $-$0.60  &	  $-$3.30   &		 &	      &  2.7   & ...  &  ...  &   disc    & Binary& 17,18\\
 HD 46703	    &	 sr*     &   $-$0.40  &	  $-$1.60   &		 &	      &        & ...  &  ...  &   disc    & Binary& 11	 \\
 HD 52961	    &	 sr*     &   	      &	  $-$4.80   &    3.00	 &  3.35      &  3.4   & ...  &  ...  &   disc    & Binary& 19   \\
 HR 4049 	    &	 pA*     &   $-$0.40  &	  $-$4.80   &		 &	      &  3.5   & ...  &  ...  &   disc    &Binary & 17	 \\
 IRAS15469$-$5311   &	 OH*     &   0.20     &	  0.00      &    1.80	 &  2.10      &  0.3   & ...  &  ...  &   disc    &Binary & 12	 \\
 IRAS17038$-$4815   &	 pA*     &   	      &	  $-$1.50   &		 &	      &  0.3   & ...  &  ...  &   disc    &Binary & 12	 \\
 IRAS17233$-$4330   &	 pA*     &   $-$0.20  &	  $-$1.00   &    1.40	 &  1.80      &  0.7   & ...  &  ...  &   disc    &  	  & 12	 \\
 IW Car		    &	 RV*     &   0.20     &	  $-$1.00   &		 &	      &        & ...  &  ...  &   disc    &  	  & 1	 \\
 SS Gem		    &	 RV*     &   $-$0.20  &	  $-$0.90   &    2.00	 &  1.60      &  0.9   & ...  &  ...  &   non$-$IR&  	  & 5a	 \\
 ST Pup		    &	 WV*     &   $-$0.10  &	  $-$1.50   &    2.10	 &  2.00      &  1.4   & ...  &  ...  &   disc    & Binary& 6	 \\
 SX Cen		    &	 RV*     &   $-$0.30  &	  $-$1.10   &    1.50	 &  1.90      &  0.6   & ...  &  ...  &   disc    &Binary & 12	 \\
 UY CMa		    &	 sr*     &   $-$0.40  &	  $-$1.3    &    1.80	 &  2.10      &  0.7   & ...  &  ...  &   disc    &  	  & 4	 \\

\hline
   \multicolumn{12}{c}{Mildly Depleted Objects.} \\
\hline

 AC Her             &	 RV*     &   $-$0.90  &	  $-$1.40  &    0.70 	&  1.20      &  0.50 &	... &	...& disc    & Binary  & 2   \\
 AZ Sgr		    &	 RV*     &   $-$0.30  &	  $-$1.60  &		&  1.30      &       & ...  &  ... & non$-$IR&         & 4	 \\
 BT Lac		    &	 RV*     &   $-$0.10  &	  $-$0.20  &    0.50	&	     &  0.10 &	... &	...& disc    &         & 15  \\ 
 EN Tra             &	 Ce*     &            &	  0.00     &	0.55	&  0.50      &  0.20 &	... &	...& disc    & Binary  & 20  \\
 EP Lyr		    &	 RV*     &   $-$0.90  &	  $-$1.80  &    1.30	&  1.40      &  1.10 &	... &	...& disc    &         & 5b  \\
 EQ Cas		    &	 RV*     &   $-$0.30  &	  $-$0.80  &    1.00	&  1.00      &  0.50 &	... &	...& non$-$IR&         & 4   \\
 IRAS09144$-$4933   &	 RV*     &   0.00     &	  $-$0.30  &		&  1.30      &       &	... &	...& disc    & Binary  & 12  \\
 IRAS16230$-$3410   &	 pA*     &   $-$0.40  &	  $-$0.70  &    1.00	&  1.10      &  0.30 &	... &	...& disc    & Binary  & 12  \\
 LR Sco		    &	 sr*     &   0.20     &	  0.00     &    0.80	&  0.60      &  0.20 &	... &	...& disc    &         & 12  \\
 QY Sge		    &	 sr*     &   $-$0.20  &	  $-$0.30  &    1.20	&  1.20      &  0.10 &	... &	...& disc    &         & 12  \\
 R Sge		    &	 RV*     &   0.10     &	  $-$0.50  &    1.10	&  1.70      &  0.30 &	... &	...& disc    &         & 5b  \\
 RU Cen		    &	 RV*     &   $-$1.10  &	  $-$1.90  &    1.00	&  1.30      &  0.90 &	... &	...& disc    & Binary  & 12  \\
 RV Tau		    &	 RV*     &   $-$0.40  &	  $-$0.40  &    0.50	&	     &  0.40 &	... &	...& disc    &         & 3   \\
 SU Gem		    &	 RV*     &   0.00     &	  $-$0.25  &    0.52	&  0.82      &  0.10 &	... &	...& disc    &         & 15  \\
 UY Ara		    &	 RV*     &   $-$0.30  &	  $-$1.00  &		&	     &  0.70 &	... &	...& disc    &         & 3   \\
 V390 Vel	    &	 RV*     &   0.00     &	  $-$0.30  &    0.90	&  1.00      &  0.40 &	... &	...& disc    &  Binary & 12  \\

\hline
   \multicolumn{12}{c}{Not Depleted  Objects.} \\
\hline

 AI Sco		    &	 RV*     &  $-$0.70 &           &	0.30	   &  0.83      &  0.10	   &   ... & ...   &  disc     &      & 4	 \\
 AR Sgr		    &	 RV*     &  	    &  $-$1.33  &	0.01	   &  0.39      &  0.10	   &   ... & ...   &  non$-$IR &      & 4	 \\
 BD$+$46 442	    &	 *       &  	    &  $-$0.79  &	$-$0.24    & $-$0.40	&  $-$0.08 &   ... & ...   &  disc     & Binary& 14	  \\
 BT Lib		    &	 RV*     &  $-$1.10 &  $-$1.10  &		   &		&  0.00    &   ... & ...   &  non$-$IR &       & 3	  \\
 DF Cyg             &	 RV*     &          &	0.00    &    $-$0.74	   &	        &  $-$0.62 &	...&	...& uncertain &        & 4       \\
 DS Aqr             &	 RV*     &          &  $-$1.10  &                  &            &          &       &       &  non$-$IR &       & 3        \\
 DY Aql             &	 RV*     &  	    &  $-$1.00  &	           &		&           &   ... & ...   &  non$-$IR&       & 12	  \\
 GP Cha 	    &	 RV*     &  $-$0.60 &  $-$0.60  &	 0.10	   &  0.30	&  $-$0.10 &   ... & ...   &  disc     &       & 12	  \\
 HD 108015	    &	 sr*     &  	    &  $-$0.09  &	0.10       & 0.11	&  $-$0.07 &   ... & ...   &  disc     & Binary& 20	  \\
IRAS06165$+$3158    &	 IR*     &  	    &  $-$0.93  &	$-$0.03    &		&  $-$0.06 &   ... & ...   &  disc     &       & 15	  \\
IRAS09060$-$2807    &	 sr*     &  $-$0.70 &  $-$0.70  &	 0.20	   &  0.00	&  0.10    &   ... & ...   &  disc     &       & 12	  \\
IRAS19157$-$0247    &	 pA*     &  0.10    &  0.10     &		   &  0.40	&	   &   ... & ...   &  disc     &Binary & 12	  \\
 R Sct		    &	 RV*     &  $-$0.20 &  $-$0.40  &	 0.10	   &		&  0.10     &   ... & ...   &  uncertain&       & 3	  \\
 RX Cap		    &	 RV*     &  $-$0.60 &  $-$0.80  &	0.00       &  0.00	&  0.20     &   ... & ...   &  non$-$IR&       & 4	  \\
 SAO 173329 	    &	 Ir*     &  	    &  $-$0.92  &	$-$0.14    & 0.41	&  0.05     &   ... & ...   &  disc    & Binary& 21	  \\
 TT Oph		    &	 RV*     &  $-$0.80 &  $-$0.80  &	 0.00	   &  0.80	&  0.00     &   ... & ...   &  non$-$IR&       & 3	  \\
 TW Cam		    &	 RV*     &  $-$0.40 &  $-$0.50  &       0.30       &  0.70      &  0.10     &	...&...     & disc     &       & 3	  \\
 TX Oph		    &	 RV*     &  $-$1.10 &  $-$1.2   &	0.00       &  0.56	&  $-$0.01  &   ... & ...   &  non$-$IR&       & 4	  \\
 U Mon		    &	 RV*     &  $-$0.50 &  $-$0.80  &	 0.00	   &  0.50	&  0.20     &   ... & ...   &  disc    & Binary& 3	  \\
 UZ Oph		    &	 RV*     &  $-$0.80 &  $-$0.70  &	 0.30	   &  0.55	&  0.10     &   ... & ...   &  non$-$IR&       & 4	  \\
 V Vul		    &	 RV*     &  0.10    &  $-$0.40  &	$-$0.13    &  0.71	&  0.10     &   ... & ...   &  disc    &       & 4	  \\
 V360 Cyg	    &	 RV*     &  $-$1.30 &  $-$1.40  &	$-$0.10    &  0.40	&  0.00     &   ... & ...   &  non$-$IR&       & 2	  \\
 V453 Oph	    &	 RV*     &  $-$2.20 &  $-$2.20  &	$-$0.03    &		&  0.40     &   ... & ...   &  non$-$IR&       & 2	  \\
 V820 Cen	    &	 RV*     &  	    &  $-$2.28  &	$-$0.07    &		&  0.40     &   ... & ...   &  non$-$IR&       & 15	  \\

\hline
   \multicolumn{12}{c}{Not Depleted  Objects (C and s$-$process Enhanced Objects).} \\
\hline
 
HD 187885	    &	 pA*  &   $-$0.60&  ...  &  ...  & ... &  ...  &   1.00	    &	1.10	  &   shell&         & 7	\\
HD 235858	    &	 pA*  &   $-$0.82&  ...  &  ...  & ... &  ...  &  	    &	1.99	  &   shell&         & 13	\\
HD 56126	    &	 sr*  &   $-$1.00&  ...  &  ...  & ... &  ...  &   1.10	    &	1.50	  &   shell&         & 7	\\
IRAS Z02229$+$6208  &	 pA*  &   $-$0.50&  ...  &  ...  & ... &  ...  &   0.80	    &	1.40	  &   shell&         & 16	\\
IRAS04296$+$3429    &	 pA*  &   $-$0.60&  ...  &  ...  & ... &  ...  &   0.80	    &	1.50	  &   shell&         & 7	\\
IRAS05113$+$1347    &	 pA*  &   $-$0.75&  ...  &  ...  & ... &  ...  &   1.09	    &	2.00	  &   shell&         & 13	\\
IRAS05341$+$0852    &	 pA*  &   $-$0.80&  ...  &  ...  & ... &  ...  &   1.00	    &	2.20	  &   shell&         & 15	\\
IRAS06530$-$0213    &	 pA*  &   $-$0.90&  ...  &  ...  & ... &  ...  &   1.30	    &	1.90	  &   shell&         & 8	\\
IRAS07430$+$1115    &	 pA*  &   $-$0.50&  ...  &  ...  & ... &  ...  &   0.60	    &	1.60	  &   shell&         & 16	\\
IRAS20000$+$3239    &	 pA*  &   $-$1.40&  ...  &  ...  & ... &  ...  &   1.70	    &	1.40	  &   shell&         & 9	\\
IRAS23304$+$6147    &	 pA*  &   $-$0.81&  ...  &  ...  & ... &  ...  &   0.91	    &	1.60	  &   shell&         & 7	\\
V448 Lac	    &	 sr*  &   $-$0.30&       &       &     &       &   0.30	    &	0.90	  &   shell&         & 7	\\

\hline					        						   	  	  
\end{tabular}
\begin{flushleft}
(1):\cite{giridhar94}; (2):\cite{giridhar98}; (3):\cite{giridhar00}; (4):\cite{giridhar05};
(5a):\cite{gonzalez97a}; (5b):\cite{gonzalez97b}; (6):\cite{gonzalez96}; (7):\cite{vanwinckel00}; (8):\cite{hrivnak03}; (9):\cite{klochkova06};
(10):\cite{kodaira70}; (11):\cite{luck84}; (12):\cite{maas05}; (13):\cite{reddy02}; (14):\cite{gorlova12} (15):\cite{rao14}; (16):\cite{reddy99}; 
(17):\cite{vanwinckel95}; (18):\cite{waelkens96}; (19):\cite{waelkens91B}; (20):\cite{vanwinckel97};(21):\cite{rao12};(22):\cite{kiss07}

\end{flushleft}
\label{depletiontable}
\end{table*}

\section{Discussion and Conclusions}

In this paper we compared RV Tauri stars with a reference sample of
post-AGB objects concerning their infrared properties, binary nature
and chemical abundances. Firstly, we showed that the RV Tauri
stars are nicely clustered in the WISE colour-colour diagram and the
clustering was confirmed by our detailed SED determinations. There are
128 RV Tauri variables in our Galaxy and 29
(23${\%}$) of them are disc sources, 63 (50${\%}$) of them are non-IR
sources and 34 (27${\%}$) of them display an SED which we classify as
uncertain. An example of
each type of SED is presented in the Fig.~11. In this way, we stress that
not all RV Tauri stars show an IR excess.

Secondly, we investigated the relation between infrared properties and
binarity for both the RV Tauri stars and the reference post-AGB
objects. Orbital parameters have been measured for
17 post-AGB objects and 7 RV Tauri stars.
As all confirmed binaries are disc sources, we conclude that
the correlation between the properties of the SED and binarity is
obvious for both RV Tauri stars and post-AGB objects. As RV Tauri
stars with a disc SED are suspected binaries, radial velocity monitoring
experiments have been focussed on these systems so far. Indeed, all
confirmed binary RV Tauri stars are disc sources. We conclude that
binarity is a widespread phenomenon among the RV
Tauri class of objects with discs but that only a minority
of them have their orbital elements determined. Homogenous radial velocity
measurements and CO observations are needed to come to general
conclusions about the binary nature of RV Tauri stars as a whole.

Finally, we reviewed the chemical studies of both the RV Tauri stars and
the reference post-AGB objects in the literature. We showed that for
both groups the
relation between infrared properties and chemical anomalies is 
complex. Strongly and mildly depleted objects are
mostly disc sources but not exclusively so. 48\% of the chemically
studied reference post-AGB objects is depleted and they all display a clear
broad band SED. The exception for the reference sample is
BD+39$^{\circ}$4926 which shows a small excess that has only been
detected thanks to the sensitivity of the WISE
experiment. Depletion is more widely observed in the RV Tauri
stars. 56\% of the chemically studied RV Tauri stars is depleted but
not all of them are confirmed disc sources. Two of the depleted RV Tauri stars
do not display a clear IR excess, but this seems to be exceptional. 
All the confirmed binaries
are disc sources but not all of them are depleted. Thus, a disc is
clearly not a sufficient condition for the dust-gas separation to take
place but it does seem to be a prerequisite.

We conclude that (1) there are disc sources which do not show depletion
patterns and the other way around (2) there are depleted sources
without a dust excess. For the first case, the suggested scenario
\citep{giridhar05,venn14} is that a deep convective envelope may prevent
the depletion signature to be observed. The main problem for this scenario is
that it is unknown how strong the dilution is. For the
latter case, the absence of a current infrared excess doesn't mean
that such a disc was not present in a earlier evolutionary phase. 
The dust-gas separation operates
on a certain timescale, thus, the dust reservoir may have already
been exhausted while we can still detect the photospheric depletion
pattern. The IR lifetime of a disc source will not only depend on
the evolution of the central object, but it will also depend on the
the evolution of the dusty disc itself. 

Several detailed radiative transfer studies, some of which were constrained
by high-spatial resolution interferometric observables,
have shown that the dusty discs can be well modelled assuming as a
passive, gas- and dust-rich irradiated disc
\citep{dominik03,hillen14}. The evolution of these discs is not well
documented yet and neither are the evolutionary processes themselves, nor the
associated timescales.  But the evolution of the
disc will imply that the infrared position of these systems in the
WISE colour-colour diagram will change accordingly.

\begin{figure}
\begin{center}
\includegraphics[bb= 0 0 1200 800, width=9cm,angle=0]{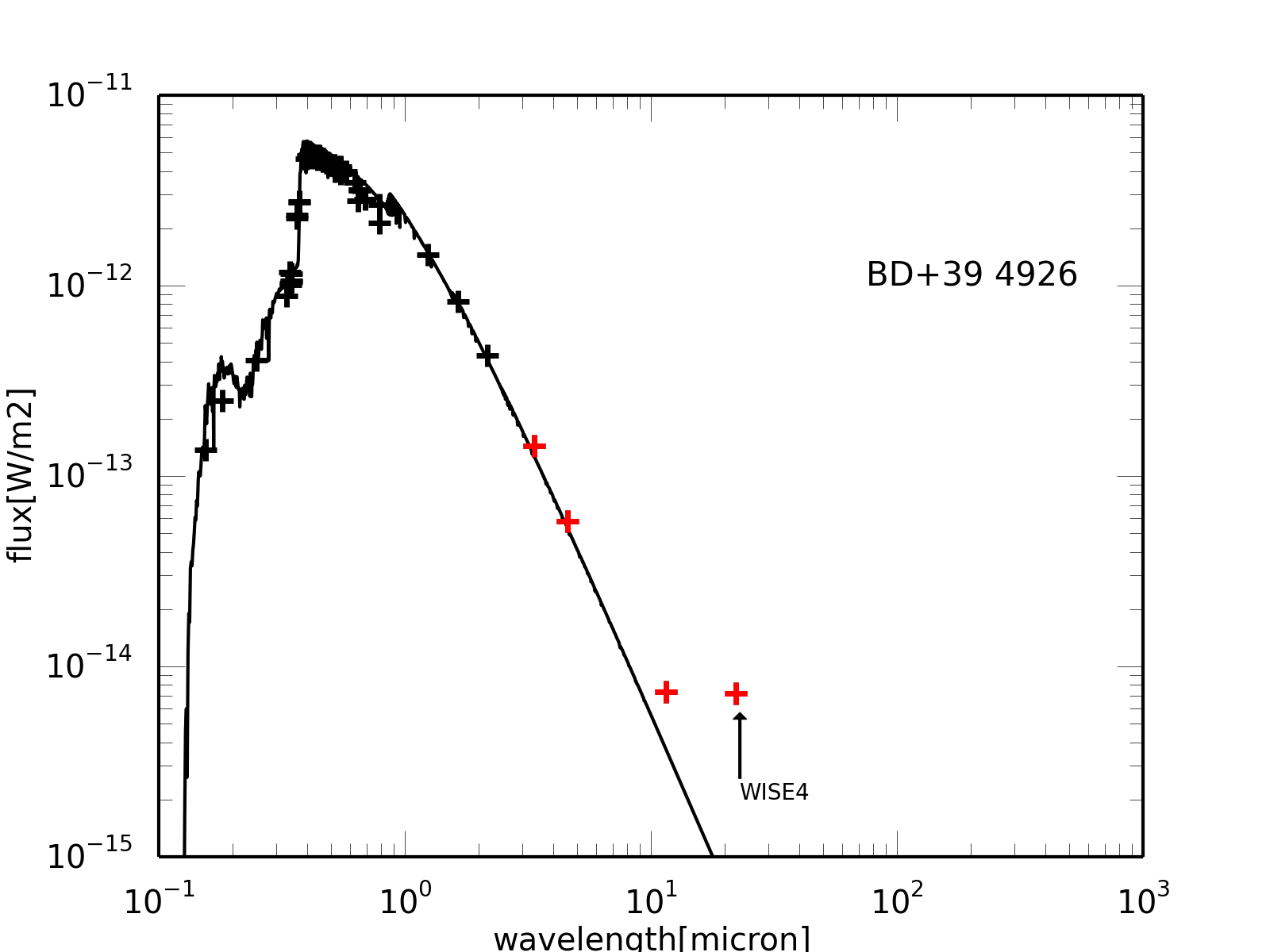}
\caption{Spectral Energy Distribution of BD$+$39${^\circ}$4926.}  
\end{center}
\end{figure}

The discs are subject to specific physical processes like grain growth,
crystallisation \citep[e.g.][]{gielen11}, and
gas evaporation \citep{bujarrabal13b}. It is likely
that the infrared luminosity of the disc will gradually decline and
will resemble the infrared properties of gas-poor debris discs. The
depletion of the affected photospheres will, however, be visible much
longer. The binary BD+39$^{\circ}$4926 may represent such a case. 
The object has been long known to present chemical peculiarities \citep{kodaira70}. By now
it is well characterised that this is due to a strong depletion of
only the refractory elements. As noted also by \citep{venn14} the
detection of a small but significant IR excess is only recent and
limited to the detection of an excess only in the WISE W4 band
(Fig.~12). The binary motion and a first estimate of the orbital period 
were based on data with a poor radial velocity accuracy and
a poor sampling \citep{kodaira70}. We therefore
included this object in our long-term radial velocity monitoring
programme \citep{vanwinckel10} using our HERMES spectrograph
\citep{raskin11}. The orbital motion is clearly detected and in
Fig.~13 we show the folded radial velocity data together with a fit
representing the Keplerian solution. In Table~\ref{BDtable} we present the full
orbital elements of BD$+$39$^{\circ}$4926 and the associated 1
$\sigma$ uncertainties. These were obtained by a Monte-Carlo analyses
in which we varied the original data point randomly within the
specific uncertainty of the velocity points. As can be seen, also from
 Fig.~13, the orbital elements are very well determined. There is
no observable to constrain the inclination. Using a 0.6 M$_{\odot}$
primary, the mass function of 0.381 M$_{\odot}$ translates into a
minimal companion mass of 1 M$_{\odot}$ and a most probable one
(assuming an inclination of 60$^{\circ}$) of 1.3 M$_{\odot}$. We do
not detect any flux contribution from the secondary, neither do we detect
any symbiotic activity and we suspect that the companion is a
non-evolved main sequence star.

Another good recently studied example of an evolved disc is AC Her
\citep{hillen15} whose interfermetric measurements show that the inner
radius of the disc is large, the large grains have been settled to the
midplane and the gas/dust ratio is small.  Interestingly also
BD+33$^{\circ}$2642, the central star of the PN G052.7+50.7 was found
to be a depleted \citep{napiwotzki94} and recently found to be a
binary as well \citep{vanwinckel14}.

From these observables it seems that the depleted photospheres remain
chemically peculiar for a longer time than the IR excess remains
detectable.  In the context of metal poor stars and the confusion
between genuine metal poor stars \citep{venn14} and strongly depleted
stars, not only the presence of an infrared excess is important, but
we show here that the binary nature of the object also seems to be a
prerequisite for the depletion process to become active.

Our homogeneous infrared study of all Galactic RV Tauri stars shows
that disc formation and hence likely binarity is indeed a widespread
phenomenon (23${\%}$) but that also many RV\,Tauri stars exist which
do not show an IR excess (50${\%}$) or show only an uncertain excess
(27${\%}$).  

The RV\,Tauri stars in the disc box share many
observational properties of disc post-AGB stars and they are likely
similar binary objects with the only difference that the RV Tauri stars
happen to be in the population II Cepheid instability strip.  The
presence of a stable disc seems to be needed but not a sufficient
condition for the depletion process to become efficient.  More radial
velocity monitoring is needed to investigate the binary properties of
the whole disc sample.

The evolutionary status of the RV\,Tauri stars without an IR excess is
less clear. As post-AGB stars with spectral types similar to these RV\,Tauri
stars are expected to have a detectable IR-excess
\citep{vanwinckel03}, the evolutionary nature of these objects remain
to be clarified. 

Many of the RV\,Tauri stars which are labelled {\sl uncertain} have just 
one fluxpoint in excess which may indicate the
presence of an expanding shell or evolved disc, but more IR data is needed to
test this. If confirmed expanding shells, then these objects may be
similar in evolutionary nature to the shell sources among the post-AGB
reference sample. If confirmed evolved discs then these objects should
be binaries in a more evolved evolutionary stages as the disc objects.

In our following paper we will deal with a systematic
abundance analysis of RV Tauri stars covering all three different SED
types: disc SEDs, uncertain SEDs and SEDs without a dust excess. We
will futher investigate  the impact of a disc onto the chemical
diversity and study the evolutionary picture of RV\,Tauri stars via
systematic studies of their photospheric composition.

\begin{table}
\caption{Orbital elements of BD$+$39$^{\circ}$4926}
\begin{center}
\begin{tabular}{l r r}
\hline
Parameter  & Value  &  $\sigma$  \\ 
\hline

$P$ (d)              & 873.06     & 0.64 \\
$a$ $sini$ (AU)      & 1.296      & 0.004 \\
$f(m)$ (M$_{\odot}$) &  0.381     & 0.004 \\
$K$ (km/s)           & 16.16      & 0.05 \\
$e$                  & 0.032      & 0.003 \\
$\omega$ ($\degr$)   & 105        & 3 \\
$T_{0}$ (JD)         & 2 455 420. & 18\\
$\gamma$ (km/s)      & -30.66     & 0.03\\
varreduc             & 99.85      & \\
\hline                  
\end{tabular}
\end{center}
\begin{tablenotes}
\item Note. Orbital parameters with their uncertainties and the variance reduction of the orbital fit expressed in
percentage of the original variance of the raw radial velocity data. 
\end{tablenotes}
\label{BDtable}
\end{table}

\begin{figure}
\begin{center}
\includegraphics[bb= 0 0 650 400, width=9cm,angle=0]{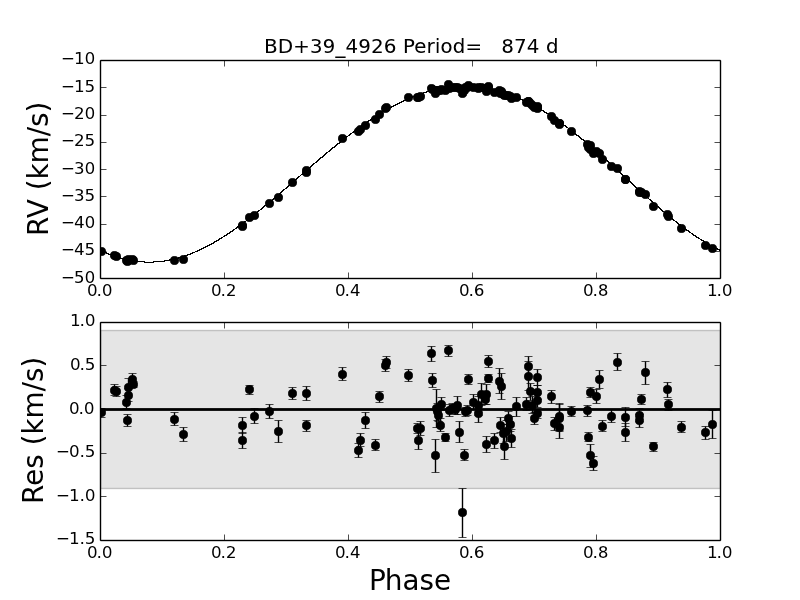}
\caption{The radial velocity curve of BD$+$39${^\circ}$4926.}
\end{center}
\end{figure}

\section*{Acknowledgments}

The authors dedicate this paper to Tom Lloyd Evans (1940$-$2014).

This work has been performed thanks to 2214-A International Research Fellowship Programme of the Scientific and Technological Research Council of Turkey (TUBITAK). 

IG would like to thank KU Leuven Astronomy Institute for their kind hospitality.

Based on observations made with the Mercator Telescope, operated on the island of La Palma by the Flemish Community, at the Spanish Observatorio del Roque de los Muchachos of the Instituto de Astrofísica de Canarias.

Based on observations obtained with the HERMES spectrograph, which is supported by the Fund for Scientific Research of Flanders (FWO), Belgium, the Research Council of K.U.Leuven, Belgium, the Fonds National de la Recherche Scientifique (F.R.S.-FNRS), Belgium, the Royal Observatory of Belgium, the Observatoire de Genève, Switzerland and the Thüringer Landessternwarte Tautenburg, Germany.   

HVW, MH, RM, DK acknlowledge support of the KU Leuven contract GOA/13/012. DK acknowledges support of the FWO grant G.OB86.13.

The following Internet-based resources
were used in research for this paper: the NASA Astrophysics Data System;
the SIMBAD database and the VizieR service operated by CDS, Strasbourg,
France.

\bibliography{mnemonic,ilklib}

\appendix

\label{lastpage}

\end{document}